\documentclass[3p, review]{elsarticle}
\usepackage{color}
\usepackage[toc]{appendix}

\usepackage{multirow}
\usepackage{color}
\usepackage{subfig}
\usepackage{url}
\usepackage{booktabs,caption,fixltx2e}
\usepackage[flushleft]{threeparttable}

\usepackage{floatrow}
\usepackage{dsfont}
\usepackage{listings}
\usepackage{amsfonts}
\usepackage{amsmath}
\usepackage{amsthm}
\usepackage{verbatim}

\newtheorem{definition}{Definition}[section]

\newtheorem{proposition}{Proposition}[section]
\newtheorem{ex}{Example}[section]

\def\eqd{\buildrel {\rm \mathfrak{D}} \over =}

\def\cw#1 { \overset{\mathbb{P}}{\underset{#1}{\longrightarrow}} }
\def\Real{\mathbb{R}}

\def\E#1{{\mathbb E}\left[#1\right]}

\def \hix#1{ {\rm HIX}\left(#1\right) }
\def \rhix#1{ {\rm RHIX}\left(#1\right) }
\def \rhix#1{ {{\rm RHIX}}\left(#1\right) }
\def \srhix#1#2{ {{\rm RHIX}_{#1}}\left(#2\right) }

\def \cix#1{ {{\rm CIX}}\left(#1\right) }
\def \rcov#1#2 {{\rm cov}_{#1}\left( #2\right)}

\begin{document}
\begin{frontmatter}
\title{Analyzing Herd Behavior in Global Stock Markets: An Intercontinental Comparison}

\author[KO]{Changki Kim}

\author[IH]{Woojoo Lee}

\author[HY]{Yang Ho Choi}

\author[EW]{Jae Youn Ahn\corref{cor2}}
\ead{jaeyahn@ewha.ac.kr}

\address[KO]{Korea University Business School, Seoul, Korea}
\address[IH]{Department of Statistics, Inha University, 235 Yonghyun-Dong, Nam-Gu, Incheon 402-751, Korea.}
\address[HY]{Department of Actuarial Science, Hanyang University, Ansan, Kyeonggi-do, Korea.}
\address[EW]{Department of Statistics, Ewha Womans University, 11-1 Daehyun-Dong, Seodaemun-Gu, Seoul 120-750, Korea.}

\cortext[cor2]{Corresponding Author}

\begin{abstract}

Herd behavior is an important economic phenomenon, especially in the context of the recent financial crises. In this paper, herd behavior in global stock markets is investigated with a focus on intercontinental comparison. 
Since most existing herd behavior indices do not provide a comparative method, we propose a new herd behavior index and demonstrate its desirable properties through simple theoretical models. As for empirical analysis, 
we use global stock market data from Morgan Stanley Capital International to study herd behavior especially during periods of financial crises in detail.

\end{abstract}


\end{frontmatter}

\vfill

\pagebreak

\section{Introduction}

Herd behavior is a term often used in financial literature to describe the comovement of members in a group without a planned direction. 
A study about herd behavior was conducted among a number of researchers over several years. 
Here we take a quick look at some studies regarding occupation or market related herding. 
\citet{Das} provide a theoretical reasoning which explains the dichotomy condition of positive and negative effect on short and long-term return respectively by institutional herding. 
Career-concerned institutional investors are deemed to prefer trading with a security dealer who is endowed with market power- further they have tendencies of repeating their past investment pattern. 
\citet{Sias} shows that institutional investors' preferences on stocks are highly correlated to those that they have invested in the past. 
It is concluded that herding exists, in which institutional investors reference other investors' trading when they invest. On the basis of the Portuguese stock market data, \citet{Holmes}
have reached to conclusion that institutional herding arises from analysts' intentional (not spurious) purpose. Such herding is window dressing for performance measurement by institutional investors. 
\citet{Walter} portray the existence of spurious herding from mutual funds in Germany, which is generated through the process of changing the index composition by fund managers. It is also proven that the actual level of herding is not high (5.11\%) 
and the herding phenomenon is not correlated with stock stabilities. \citet{Hirsh} review previous studies in regard to four things in the capital market: herd behavior, interactions between payoff and reputational, social learning and informational cascades. 
They evaluate these studies from the perspective of explanatory power and analyze the incentives of many various behaviors including herding.

According to \citet{Kim}, a company with a complexity of structure induces more herding due to the the securities analysts' difficulty of analyzing it. 
It was concluded that analysts have herding behavior, in which industrially or geographically diversified companies are evaluated as lower valued of companies compared to other undiversified companies. 
\citet{Jega} analyze various herding behaviors of sell-side analysts. From the analysis, a sharper reaction of the market to revised recommendations away from the consensus is induced as analysts' herding is recognized 
by market participants. Using the modeling, \citet{NelsonL} analyzes herd behavior, which occurs through the decision making by Initial Public Offering (IPO) companies. It is induced that herding does not last 
long and cannot be simplified with features.

\begin{figure}
\includegraphics[width=0.6\linewidth]{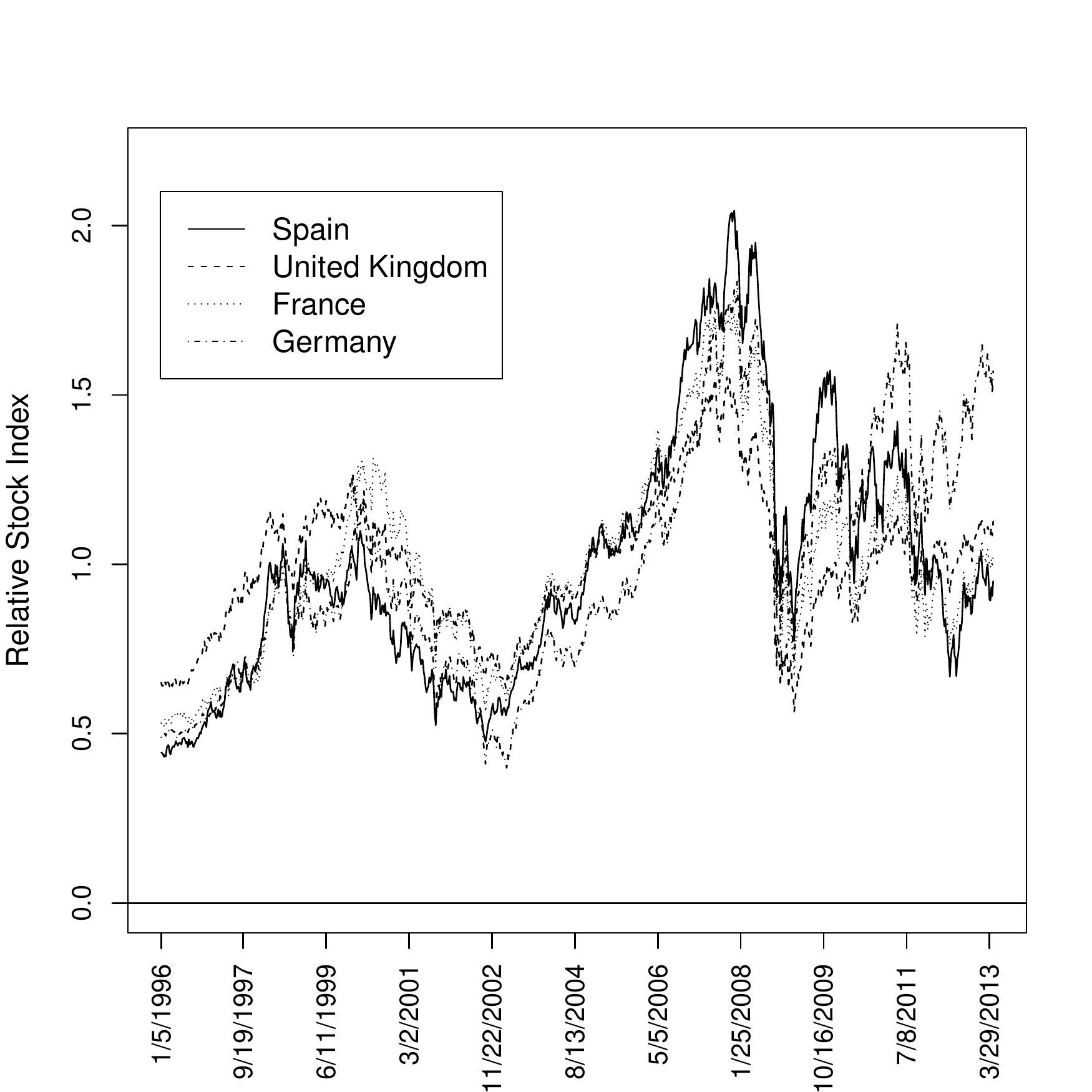}
         \centering\caption{Stock Index: Selected European Countries}\label{fig0}
\end{figure}

Herding is highlighted as an important economic phenomenon, especially in relation to the recent financial crises. \citet{Sushil} summarized the meaning of herd behavior, the reasons for herding, and some examples from research that identify special herding occurrences. As shown in Figure \ref{fig0}, which shows the daily stock 
indices for some selected countries in Europe from 1996 to April 2013, the movement of a representative stock index of one country almost imitates that of other countries' stock indices. The tendency of comovement in European stock markets appears to be strong, perhaps because 
all countries in European countries are in continual contact with one another. 
Based on this observation, we ask several questions. Are there comovements, such as those in Europe, in the stock markets of other continents? 
If so, can we identify herd behavior? How do we measure such herd behavior? These questions lead us to the study of how to measure and compare the comovement of stock prices among continents.

In this paper, we discuss herd behavior in global stock markets and also make an intercontinental comparison. We focus particularly on periods of financial crisis: the Asian financial crisis beginning in July 1997, the dot com bubble burst starting in early 2000, the global financial crisis beginning in 2008, and the ongoing European sovereign debt crisis. 
We would like to to develop a herd behavior index not only to measure the degree of herding in a subject but also to make comparisons between the herd behaviors of different subjects. However, because most existing herd behavior indices do not provide a comparative method, another goal of this research is to develop a normalized herd behavior index.

Several measures have been used by researchers to explain herd behavior. \citet{Chris} introduced the cross-sectional standard deviation (CSSD) of returns, i.e.,
the ratio of the average proximity of individual asset returns to the realized market average, in order to identify herd behavior. \citet{Chang}, \citet{Chiang}, and \citet{Prosad} used CSSD
to show that several countries demonstrated strong evidence of herd behavior. However, their results did not indicate the actual timing of the occurrence of herd behavior. The second one is the implied correlation index (CIX), suggested by \citet{Skintzi}, which is defined as the ratio of the sum of the weighted covariance to that of the weighted variance between stocks. In order to have more accurate measure of future correlation, CIX is computed using the implied volatility from the portfolio option and the implied volatilities from options on each of the portfolio assets. All implied volatilities should be derived from options with the same maturity date. Since, however, this index is a 
kind of generalized correlation, it has a certain limit for explaining herd behavior as will be explained below. 
Lastly, the Herd Behavior Index (HIX) suggested by \citet{Dhaene2} is defined as the ratio of the estimate of the variance of the real market index to that under extreme herd behavior. 
Although HIX is an index that can spot herding effects, it may not suit our purpose, which is an intercontinental comparison of herd behaviors.

As shown in Figure \ref{fig.cix}, CIX misleads researchers into believing that herd behavior is not observable in Case 1 under a high correlation between two stocks. 
From Figure \ref{fig1}, one can also observe that the market with one large stock always has a large HIX. 
More specifically, CIX and HIX may suffer from a normalization problem, which implies that HIX with the same value may have a different interpretation as to the degree of herd behavior under the different settings.
Therefore in this paper, we propose a new herd index: the
revised herd behavior index (RHIX), which can be interpreted as the normalized version of CIX and HIX. 
By applying this new herd behavior index, we analyze herd behavior in the financial market across continents.

As shown in \citet{Dhaene2}, if the prices of the underlying assets, such as vanilla option prices, are available, the distribution of the underlying assets can be found; then the implied version of HIX can be calculated.
Hence HIX can be used as a forward looking heard behavior index. 
While the same method can be used for the calculation of the implied version of RHIX, we need the prices of options based on a continent's index, which is not available. 
Furthermore because our purpose is not a prediction but a historical review, we use historical stock indices to calculate RHIX directly. 
In the actual data analysis in Section \ref{section.data}, weekly stock price data are used to estimate RHIX, and for the calculation of confidence interval, 
we assume that the stock indices follow the lognormal distribution.

This paper is organized as follows. In Section \ref{section.2}, we briefly review the concept of herd behavior indices; CIX and HIX. 
Then, we define a new herd behavior index, RHIX, and discuss its characteristics and differences from CIX and HIX. 
In Section 3, we compute RHIX using the global stock market data from 
Morgan Stanley Capital International (MSCI), one of the leading providers of investment decision support tools to investment institutions, 
and analyze herd behavior in global stock markets. Finally Section 4 concludes the paper with suggestions on the direction of future research.

\section{Herd Behavior Indices}\label{section.2}

In this section, we demonstrate HIX and investigate its limitations for our research purpose, the intercontinental comparison of herd behavior. To overcome these limitations, we propose a revised version of HIX. This new herd behavior index is called the revised herd behavior index (RHIX). It will be shown that RHIX has various desirable properties.

To basic notations are introduced to describe herd behavior indices in a formal way. Consider $d$ different stock indices from $d$ different countries. Let ${\bf X}(t)=(X_1(t),\cdots, X_d(t))$
where $X_i(t)$ is the representative stock index of the $i$-th country at time $t$. The weight vector $\bf w$ is denoted by
\[
{\bf w}=(w_1, w_2,\cdots, w_d),\;\;\text{with}\;\;  w_i \ge 0 \quad \hbox{for any} \;\; i \in \left\{1, 2, \cdots, d \right\}.
\]
For a given $\bf w$ and ${\bf X}(t)$, consider a new weighted market index, $S(t)$, which is defined as a linear combination of $d$ different stock indices:
\[
S(t)=\sum\limits_{i=1}^{d}w_iX_i(t).
\]
We define the joint distribution of ${\bf X}(t)$ as
$$F_{{\bf X}(t)}({\bf x})={P}({X_1(t)\le x_1,\cdots, X_d(t)\le x_d}),$$
and let
$$F_{X_i(t)}(x_i)={P}(X_i(t)\leq x_i),$$
be the marginal distribution of $X_i(t)$, $i\in\{1,\cdots, d\}$. Further, let $U$ be a uniform random variable on $(0,1)$.
In addition, we use $F^{-1}_{X_i(t)}$, $\sigma_{X_i(t)}$ and $\mu_{X_i(t)}$ as the inverse of $F_{X_i(t)}$, standard deviation of $X_i(t)$ and expectation of $X_i(t)$, respectively.

Intuitively, herd behavior can be represented in terms of positive dependence between the components of a random vector. As an extension of this context, the strongest herd behavior corresponds to perfect positive dependence. This concept, formalized as comonotonicity, 
has been developed in the realms of finance and insurance research for more than 20 years. A formal definition and the basic properties of comonotonicity can be found in the Appendix. The mathematical properties of comonotonicity are summarized in \citet{Dhaene}, \citet{Cheung}, and \citet{Nam}.
In particular, as shown in Theorem {A.1} in the Appendix, \citet{Dhaene} showed that
${\bf X}(t)$ is comonotonic if and only if
$$ {\bf X}(t)\eqd \left(F_{X_1(t)}^{-1}(U), \cdots, F_{X_d(t)}^{-1}(U) \right),$$ where $U$ is a uniform (0,1) random variable and $\eqd$ means equal in the distribution.
The second expression, which will be used as a definition of comonotonicity throughout this paper, shows that comonotonic risks are driven by a single risk factor $U$. The increase of $X_{i}$
implies the increase of all the other stock prices because $F^{-1}_{X_i(t)}(U)$ is a non-decreasing function of $U$ for all $i\in\{1,\cdots, d\}$. For convenience, we write $(X_1^c(t),\cdots, X_d^c(t))
:=\left(F_{X_1(t)}^{-1}(U), \cdots, F_{X_d(t)}^{-1}(U) \right)$ to represent the comonotonic elements of ${\bf X}(t)$. Similarly, the weighted sum of the components of the comonotonic vectors,
$S^c(t)$, is defined as
\[
S^c(t)=\sum\limits_{i=1}^{d}w_iX_i^c(t). 
\]
Brief properties of $S^c(t)$ can be found in Theorem {A.2} in the Appendix.

\subsection{CIX and HIX}\label{cix}
First, the obvious approach to describe a herd behavior is the use of pairwise correlations between stock prices.
\citet{Skintzi} defined the weighted average of the pairwise correlations among stock prices, the implied correlation index (CIX), as
\begin{equation} \label{def.cix}
\cix{{\bf w}, {\bf X}(t)}= \frac{\sum\limits_{i\neq j} w_i w_j {\rm cov}\left( X_i(t), X_j(t) \right)}  {\sum\limits_{i\neq j} w_i w_j \sigma_{X_i(t)}\sigma_{X_j(t)}},
\end{equation}
Note that CIX is exactly the same value as the correlation between two assets if $d=2$.



As specified in \citet{Dhaene2}, the use of the pairwise correlation may fail to capture herd behavior. To overcome this limitation of CIX as a herd behavior index, the new degree of comovement of stock prices, the HIX measures, is defined as
\begin{equation} \label{def.hix2}
\hix{{\bf w}, {\bf X}(t)}= \frac{{\rm Var}[S(t)]}{{\rm Var}[S^c(t)]}.
\end{equation}
which is the ratio of the variance of the real market situation to that under the comonotonic assumption. $Var(S^c(t))$ in \eqref{def.hix2} represents the variance of the weighted market index
under the comonotonic assumption. Note that CIX in \eqref{def.cix} and HIX in \eqref{def.hix2} can be calculated from the vanilla option prices as specified in \citet{Skintzi} and \citet{Dhaene2}.

The following proposition states some properties of CIX. Note that the condition in \eqref{cix.low.1} can be interpreted as a condition in complete mixability. 
Information of complete mixability can be found in \citet{Ruschendorf}, \citet{Ruodu} and \citet{Ruodu2}. 

\begin{proposition}\label{prop.cix}
Let $\E{X_i(t)}=\mu_{X_i(t)}$, ${\rm Var}[X_i(t)]=\sigma_{X_i(t)}^2$, and $w_i$ be given for $i=1, \cdots, d$.
 \begin{enumerate}
  \item[1.] For any ${\bf X}(t)$, we have
  \begin{equation}\label{prop.cix.eq1}
-\frac{\sum\limits_{i}w_i^2\sigma_{X_i(t)}^2 }{ \sum\limits_{i\neq j}w_i w_j \sigma_{X_i(t)}\sigma_{X_j(t)} } \le \cix{{\bf w}, {\bf X}(t)} \le
\frac{\sum\limits_{i\neq j}w_i w_j {\rm cov}\left[X_i^c(t),X_j^c(t) \right] }{ \sum\limits_{i\neq j}w_i w_j \sigma_{X_i(t)}\sigma_{X_j(t)} }.
  \end{equation}
  \item[2.] The lower bound of \eqref{prop.cix.eq1} is attained if and only if
  \begin{equation}\label{cix.low.1}
    P\left(\sum\limits_{i=1}^{d}w_i X_i(t)=c\right)=1,
  \end{equation}
for some constant $c\in\Real$.
  \item[3.] The upper bound of \eqref{prop.cix.eq1} is attained if and only if ${\bf X}(t)$ is comonotonic.
  \item[4.] If elements of ${\bf X}(t)$ are pairwise uncorrelated, then $\cix{{\bf w}, {\bf X}(t)}=0$.
 \end{enumerate}
\end{proposition}

HIX in \eqref{def.hix2} satisfies $$0\le \hix{{\bf w}, {\bf X}(t)}\le 1$$
because the variance is nonnegative and ${\rm Var}[S(t)]\leq {\rm Var}[S^c(t)]$.
Furthermore, each bound can be achieved under the following conditions:
$$0=\hix{{\bf w}, {\bf X}(t)}\quad\hbox{if and only if}\quad S(t) ~\hbox{is constant}$$
and
$$\hix{{\bf w}, {\bf X}(t)}= 1 \quad\hbox{if and only if \; ${\bf X}(t)$ is comonotonic.}$$

\citet{Skintzi}  and \citet{Dhaene2} applied CIX and HIX to the Dow-Jones industrial average and interpreted the high peaks of the indices as signs of stress before crisis.
Because the indices showed a tendency to increase when the market index decreases in their empirical study, the indices are interpreted as fear or stress indicators.
However as will be explained in Section \ref{sec.limit}, CIX or HIX may not be an appropriate measure of
herd behavior for our research purpose, which is the intercontinental comparison of herd behavior.

\subsection{Limitations of CIX and HIX}\label{sec.limit}
In this section, we explain as to why CIX and HIX cannot be used for the intercontinental comparison of herd behavior. For pedagogical purposes, the multivariate geometric Brownian motion of stock indices is considered.
This model was also used in \citet{Dhaene2} in order to explain the properties of HIX.
We emphasize that the definitions of CIX and HIX do not require this model assumption. 

The return on asset $i$, ${\rm d}X_i(t)/X_i(t)$ consists of two parts. The first is $r_i {\rm d}t$, where $r_i$ is a measure of the average rate of growth of the asset $i$. The second is
represented by $\sigma_i {\rm d}B_i(t)$. Here, $\sigma_i(t)$ is the volatility of asset $i$ and ${\rm d}B_i(t)$ is the instantaneous changes in the standard Brownian motion. We use
$\rho_{ij}$ for the correlation between two random variables ${\rm d}B_i(t)$ and ${\rm d}B_j(t)$ for $i,j\in\{1, \cdots, d\}$. The corresponding model is described as
\begin{eqnarray}
\begin{cases}\label{model.1}
\displaystyle\frac{{\rm d}X_1(t)}{X_1(t)}=r_1{\rm d}t+\sigma_1{\rm d}B_1(t)\\
\;\;\;\;\;\;\;\;\;\;\;\;\;\;\;\;\vdots \\
\displaystyle\frac{{\rm d}X_d(t)}{X_d(t)}=r_d{\rm d}t+\sigma_d{\rm d}B_d(t),
\end{cases}
\end{eqnarray}
where $(B_1(t), \cdots,B_d(t))$ are the multivariate Brownian motion.

%

For explanatory purposes, we consider a model with only two assets and further define $S(t)=w_{1}X_{1}(t)+w_{2}X_{2}(t)$ for a fixed $t$. For this model, HIX is calculated as
\begin{eqnarray*}
\hix{{\bf w}, {\bf X}(t)}=\frac{w^2_{1}\sigma^2_{X_1(t)}+w^2_{2}\sigma^2_{X_2(t)}+2w_{1}w_{2}\sigma_{X_1(t)}\sigma_{X_2(t)}\, {\rm corr}(X_{1}(t),X_{2}(t))}{w^2_{1}\sigma^2_{X_1(t)}+w^2_{2}\sigma^2_{X_2(t)}+2w_{1}w_{2}\sigma_{X_1(t)}\sigma_{X_2(t)}\, {\rm corr}(F^{-1}_{X_1(t)}(U),F^{-1}_{X_2(t)}(U))},
\end{eqnarray*}
where $\E{X_{i}(t)}=e^{r_i t}$ and $\sigma^2_{X_i(t)}={X_{i}(0)}^2 e^{2r_i t}(e^{\sigma^2_{i}t}-1)$. The correlation coefficients are
\begin{equation}\label{lem.eq.10}
{\rm corr}(X_{1}(t),X_{2}(t))=\frac{\exp(\rho_{12}\sigma_{1}\sigma_{2}t)-1}{\sqrt{\exp(\sigma^2_{1}t)-1}\sqrt{\exp(\sigma^2_{2}t)-1}}
\end{equation}
and
\begin{equation}\label{lem.eq.11}
{\rm corr}(F^{-1}_{X_1(t)}(U),F^{-1}_{X_2(t)}(U))=\frac{\exp(\sigma_{1}\sigma_{2}t)-1}{\sqrt{\exp(\sigma^2_{1}t)-1}\sqrt{\exp(\sigma^2_{2}t)-1}}.
\end{equation}
Similarly, CIX can be calculated as
\begin{equation}\label{cix.eq.1}
\begin{aligned}
\cix{{\bf w}, {\bf X}(t)}  &=  \frac{2w_{1}w_{2}\sigma_{X_1(t)}\sigma_{X_2(t)}\, {\rm corr}(X_{1}(t),X_{2}(t))}{2w_{1}w_{2}\sigma_{X_1(t)}\sigma_{X_2(t)}}\\
  &= {\rm corr}(X_{1}(t),X_{2}(t))
\end{aligned},
\end{equation}
with ${\rm corr}(X_{1}(t),X_{2}(t))$ as defined in \eqref{lem.eq.10}.

Example \ref{example.big} shows some of the limitations of HIX and CIX as a herd behavior index between two different subjects with different settings. The example shows that HIX can be dominated by either the
variance or the weight of the assets regardless of the correlation between two assets. The numerical values $\sigma_{1}=0.2,r_1=r_2=0.03$ and $t=1$ will be applied.

\begin{ex}\label{example.big}
In this example, consider the following two cases.\\
\indent{\bf Case 1.} Two stocks have equal weight, $w_{1}=w_{2}$, with different variances.\\
\indent{\bf Case 2.} Two stocks have equal variance, $\sigma_{X_1(t)}^2=\sigma_{X_2(t)}^2$, with different weights.\\

\begin{figure}[!ht]
  \centering
  \includegraphics[width=.49\textwidth]{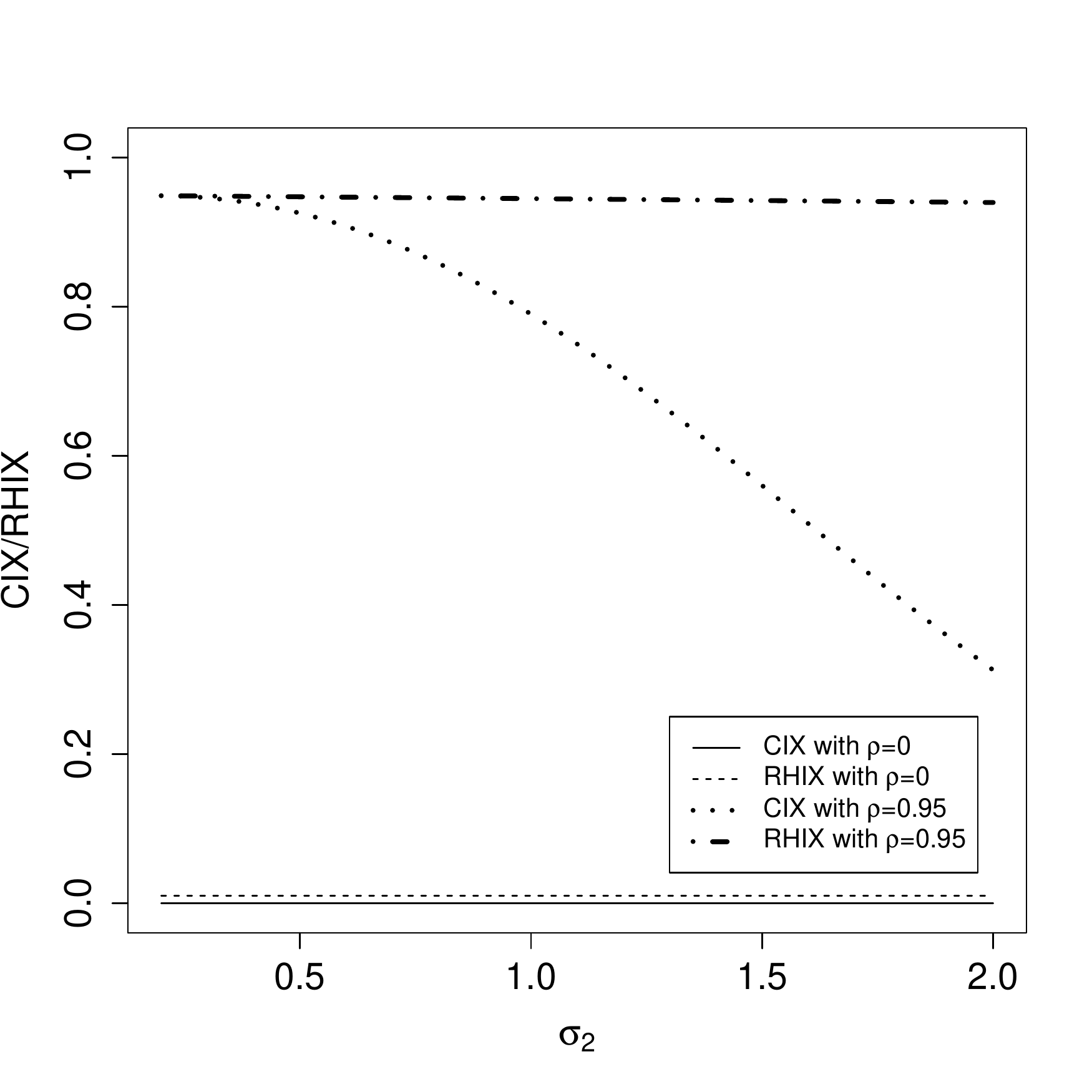}
  \caption{CIX and RHIX Comparison by Volatility Effect}\label{fig.cix}
\end{figure}

\begin{figure}[!ht]
  \centering
  \subfloat[HIX and RHIX with Volatility Effect]{\includegraphics[width=.49\textwidth]{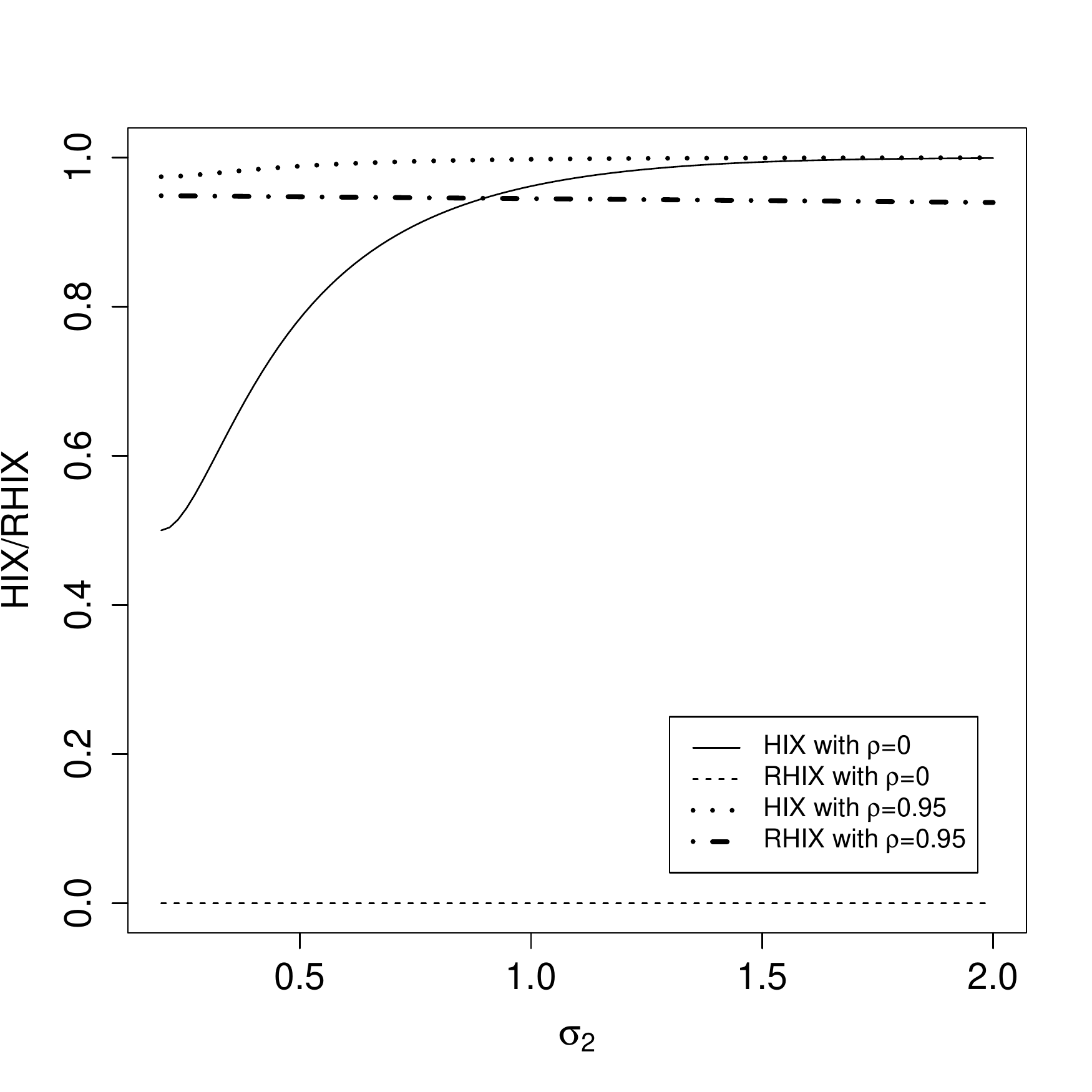}}
  \subfloat[HIX and RHIX with Weight Effect]{\includegraphics[width=.49\textwidth]{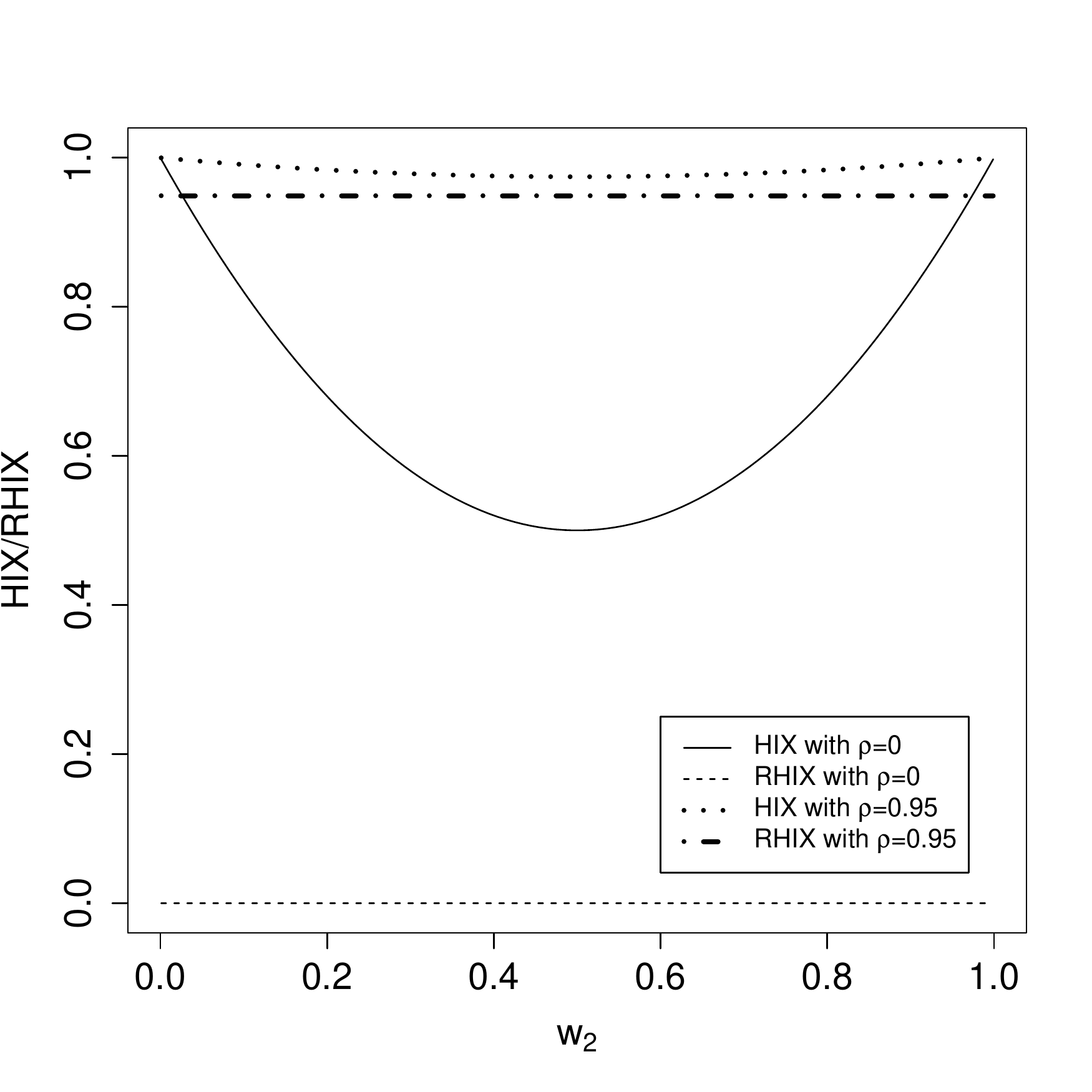}}\\
  \caption{HIX and RHIX Comparison}\label{fig1}
\end{figure}

In Case 1, we consider two extreme settings: one with a high correlation between two stocks, $\rho_{12}=0.95$, and the other with an independent assumption, $\rho_{12}=0$.
As shown in Figure \ref{fig1} (a), regardless of whether the two stocks are correlated or not, HIX increases to 1 as the variance, $\sigma_{X_2(t)}^2$, of one stock increases.
Hence when one stock price has relatively a larger variance than the other stock price, HIX may give a false impression that the two stocks are almost comonotonic even though the two stocks are uncorrelated.
Similarly, CIX is significantly decreasing as the volatility of one stock price is increasing. Hence even though the two stocks are almost comonotonic, CIX may give a false impression that the two stocks are far from comonotonic.

In Case 2, we can argue the similar logic given the same correlation settings: one with a high correlation, $\rho_{12}=0.95$, and the other with an independent assumption, $\rho_{12}=0$. As shown in 
Figure \ref{fig1} (b), regardless of whether the two stocks are correlated or not, HIX increases to 1 as the weight of one stock, $w_2$, increases. Hence, when one stock price has a much larger weight than that of the other stock, HIX may give a false impression that the two stocks are almost ``comonotonic" even though the two stocks are uncorrelated.
Finally, note that CIX does not depend on the weights of stocks for $d=2$ as specified in \eqref{cix.eq.1}.

\end{ex}

We note that rather, the abnormal behavior of CIX and HIX described in Example \ref{example.big} was not a significant issue in \citet{Skintzi} and \citet{Dhaene2}. Their goal was to measure herd behavior in a single market, where one may assume that the weights or volatility ratio of stock prices are almost constant.

\subsection{The Revised HIX}

To overcome these limitations of CIX and HIX, we propose a revised version of HIX as follows.

\begin{definition}\label{def.rhix}
Define the weighted covariance of the random vector ${\bf X}(t)$ with weight $\bf w$ as
\begin{equation}\label{eq.rhix.def}
 \rcov{\bf w}{{\bf X}(t)} :=\sum\limits_{i\neq j} w_i w_j {\rm cov}\left( X_i(t), X_j(t) \right).
\end{equation}
Using the notation in \eqref{eq.rhix.def}, we define the Revised HIX (RHIX) as
\begin{equation}\label{rhix}
 \begin{aligned}
 \rhix{{\bf w}, {\bf X}(t)} &=\frac{ \rcov{\bf w}{{\bf X}(t)}  }{  \rcov{\bf w}{{\bf X}^c(t)} } \\
 &=\frac{\sum\limits_{i\neq j} w_i w_j {\rm cov}\left( X_i(t), X_j(t) \right)}  {\sum\limits_{i\neq j} w_i w_j {\rm cov}\left( X_i^c(t), X_j^c(t) \right)},\\
\end{aligned}
\end{equation}
which is the ratio of the weighted average covariance of stock indices to the weighted average covariance of comonotonic stock indices.
\end{definition}

Similar to CIX and HIX, RHIX can also be defined using the option based expression.
However RHIX in \eqref{rhix} requires the calculation of covariance or correlation terms, whose estimation from derivative securities is painstaking as specified in \citet{Skintzi}.
We can alternatively express RHIX as
\begin{equation}\label{implied.rhix}
 \rhix{{\bf w}, {\bf X}(t)} =\frac{ {\rm var}\left[  S(t)\right]-\sum\limits_{i=1}^d w_i^2 {\rm var}\left[ X_i(t) \right] }{ {\rm var}\left[  S^c(t)\right]-\sum\limits_{i=1}^d w_i^2 {\rm var}\left[X_i(t)\right] }.
\end{equation}
Now, RHIX in \eqref{implied.rhix} does not require the estimation of the covariance terms. More specifically, we only require the estimate of portfolio and asset volatilities for the calculation of RHIX.

To compare RHIX in \eqref{rhix} with HIX, we express HIX as follows:
\begin{equation}\label{eq.hix.1}
\begin{aligned}
 \hix{{\bf w}, {\bf X}(t)}
            &= \frac{{\rm Var}\left[\sum\limits_{i=1}^{d} w_iX_i(t) \right] }{{\rm Var}\left[\sum\limits_{i=1}^{d} w_iX_i^c(t) \right]}  \\
            &=\frac{\sum\limits_{i,j}^{d} w_iw_j{\rm cov}(X_i(t), X_j(t))}{\sum\limits_{i,j}^{d} w_iw_j{\rm cov}(X_i^c(t), X_j^c(t))};
\end{aligned}
\end{equation}
We immediately note that unlike HIX, RHIX depends on the pure covariance terms only.
By removing the variance terms from both the numerator and the denominator in \eqref{eq.hix.1}, we obtain the
definition of RHIX as in \eqref{rhix}.
Further, by comparing \eqref{def.cix} and \eqref{rhix}, we also note that RHIX only differ from CIX in the numerator, where the covariance terms of comonotonic stock prices, ${\rm cov}\left[X_i^c(t), X_j^c(t) \right]$, are replaced by $1$.
Hence, a possible prejudice is that RHIX would be similar to CIX or HIX as herd behavior indices.
However, RHIX is less affected by the marginal variances or weights of individual stock indices and reflects the comovement relation more properly
as shown in Example \ref{example.big.2} below. This implies that only RHIX has the 
desirable properties for our research purpose, the comparison of the herd behavior of two markets with different settings.
The following example represents a significant difference between  RHIX and other herd behaviors, CIX and HIX, more concretely.

\begin{ex}[Revisit Example \ref{example.big}]\label{example.big.2}
First, note that the revised herd index with two assets is calculated by
\begin{equation}\label{rhix.2}
\rhix{{\bf w}, {\bf X}(t)}=\frac{corr(X_{1}(t),X_{2}(t))}{corr(F^{-1}_{X_1(t)}(U),F^{-1}_{X_2(t)}(U))}=\frac{\exp(\rho_{12}\sigma_{1}\sigma_{2}t)-1}{\exp(\sigma_{1}\sigma_{2}t)-1}.
\end{equation}
The generalized version of \eqref{rhix.2} for any $d\ge 2$ can be found in \eqref{11} in Section \ref{section.3.2}.

Again we consider Case 1 from Example \ref{example.big} with two extreme settings: one with $\rho_{12}=0.95$ and the other with $\rho_{12}=0$.
As already discussed, the interpretations of CIX and HIX are vague with varying variance of the stock price.
On the contrary, RHIX can exactly show if herd behavior is present between two stocks, as depicted in Figure \ref{fig1} (a).


Case 2 from Example \ref{example.big} is similar given two extreme settings: one with $\rho_{12}=0.95$ and the other with $\rho_{12}=0$. As the weight of one stock, $w_2$,
increases, HIX cannot distinguish whether herd behavior exists between two stocks, while RHIX can clearly show the presence of herd behavior as depicted in Figure \ref{fig1} (b).
\end{ex}

While CIX, HIX and RHIX are measures of the comovement of stock prices, RHIX is better for our purpose because it can be interpreted as a normalized CIX or HIX. As shown in Example
\ref{example.big.2}, regardless of the other parameters, RHIX is always $0$ as long as two assets are uncorrelated and is always $1$ as long as two assets are comonotonic.
This interpretation can be generalized to arbitrary number of assets. Thus, $0$ becomes the reference for the no herd behavior and $1$ becomes the reference for the perfect behavior.
On the other hand, it is difficult to set a reference for the no herd behavior with HIX and to set a reference for the perfect herd behavior with CIX. The following proposition shows the properties of RHIX.


\begin{proposition}\label{prop.rhix}
Let $\E{X_i(t)}=\mu_{X_i(t)}$, ${\rm Var}[X_i(t)]=\sigma_{X_i(t)}^2$, and $w_i$ be given for $i=1, \cdots, d$.
 \begin{enumerate}
  \item[1.] For any ${\bf X}(t)$, we have
  \begin{equation}\label{prob.rhix.eq1}
-\frac{\sum\limits_{i}w_i^2\sigma_{X_i(t)}^2 }{ \sum\limits_{i\neq j}w_i w_j {\rm cov}(X_i^c(t), X_j^c(t)) } \le \rhix{{\bf w}, {\bf X}(t)} \le 1.
  \end{equation}
  \item[2.] The lower bound of \eqref{prob.rhix.eq1} is attained if and only if
  $$P\left(\sum\limits_{i=1}^{d}w_i X_i(t)=c\right)=1,$$ for some constant $c\in\Real$.
  \item[3.] The upper bound of \eqref{prob.rhix.eq1} is attained if and only if ${\bf X}(t)$ is comonotonic.
  \item[4.] If elements of ${\bf X}(t)$ are pairwise uncorrelated, then $\rhix{{\bf w}, {\bf X}(t)}=0$.
 \end{enumerate}
\end{proposition}

The proof is given in the Appendix. Since CIX, HIX and RHIX can be used to measure the herd behaviors under different units, the invariance property under a linear transformation is desirable.
The following proposition, whose proof is in the Appendix, states that they indeed have such an invariance property.
The inner product between two vectors is defined as
\[
 {\bf a}\cdot {\bf b}=(a_1b_1, \cdots, a_db_d)\quad \hbox{for}\quad {\bf a}=(a_1, \cdots, a_d), {\bf b}=(b_1, \cdots, b_d)\in \Real^d.
\]
\begin{proposition}\label{prop.linear.1}
 Let ${\bf a}:=(a_1, \cdots, a_d)$, ${\bf a}^{-1}:=\left( \frac{1}{a_1}, \cdots, \frac{1}{a_d}\right) $ and ${\bf b}:=\left( b_1, \cdots, b_d\right) $ with $a_i \neq 0$ for any $i\in\{1, \cdots, d\}$.
 \begin{enumerate}
  \item[1.] CIX, HIX and RHIX are invariant under linear transformation, i.e.,
  \[\begin{cases}
     \cix{  {\bf w},\alpha {\bf X}(t)+{\bf b} }=\cix{{\bf w}, {\bf X}(t)},\\
     \hix{  {\bf w},\alpha {\bf X}(t)+{\bf b} }=\hix{{\bf w}, {\bf X}(t)},\\
     \rhix{  {\bf w},\alpha {\bf X}(t)+{\bf b} }=\rhix{{\bf w}, {\bf X}(t)},\\
    \end{cases}
  \]
  for $\alpha\neq 0$.
  \item[2.] The following equalities holds
 \begin{equation}\label{linear.eq.1}
 \begin{cases}
  \cix{{\bf a} \cdot {\bf w}, {\bf a}^{-1}\cdot{\bf X}(t)} =\cix{{\bf w}, {\bf X}(t)}.\\
  \hix{{\bf a} \cdot {\bf w}, {\bf a}^{-1}\cdot{\bf X}(t)} =\hix{{\bf w}, {\bf X}(t)}.\\
  \rhix{{\bf a} \cdot {\bf w}, {\bf a}^{-1}\cdot{\bf X}(t)} =\rhix{{\bf w}, {\bf X}(t)}.
 \end{cases}
 \end{equation}
 \end{enumerate}
\end{proposition}

\section{Numerical Illustration}\label{section.data}
\subsection{Data}
We describe the nature of the data used for the numerical analysis. The numbers in the brackets at the right of each country represent the aggregate value of the stock market in trillions of US dollars (at the last week of April in 2013). We use the available data between the first week of 1996 and the last week of April in 2013.

We analyze the representative stock indices from 11 Asian countries: Australia ($1.185$), China ($5.109$), Hong Kong ($1.659$), India ($1.241$), Indonesia ($0.051$), Japan ($4.542$),
Malaysia ($0.488$), the Philippines ($0.256$), South Korea ($1.134$), Taiwan ($0.866$), and Thailand ($0.477$). 
Further, we also use the currency data in the countries to calculate the dollar values of the aggregate value of the stock markets. 
Note that in Section \ref{section.3.2}, we estimate the RHIX of Asia with and without China in order to study the effect of China's stock index on the RHIX of Asia.

In Europe, we observe a total of nine countries: Belgium ($0.262$), Denmark ($0.261$), England ($0.644$), France ($1.654$), Germany ($3.222$), the Netherlands ($0.325$), Norway ($1.894$), Spain ($1.683$), and Switzerland ($0.806$). In South America, we observe Argentina ($0.045$), Brazil ($1.250$), Chile ($0.328$), and Venezuela ($0.021$). Stock prices in Brazil and Chile account for over $95\%$ of the aggregate market value. In North America, we observe Canada ($1.90$), Mexico ($0.003$), and the USA ($20.5$). Stock prices in Canada and the USA account for over $99\%$ of the aggregate market value.

For Africa and the Middle East, we cannot collect data for the entire period. In Africa, we observe Egypt ($0.052$), Mauritius ($0.011$), Morocco ($0.052$), and South Africa ($0.500$) from the first week of July 1998 to the last week of April 2013. More than $80\%$ of the money is concentrated in South Africa. The data on the Egyptian market is not available during the Egyptian Revolution period from late January 2011 to late March 2011. In the Middle East, we observe Jordan ($0.026$), Oman ($0.022$), Qatar ($0.132$), and Saudi Arabia ($0.384$) from the first week of July 2000 to the last week of April 2013. Almost $70\%$ of the money is concentrated in Saudi Arabia.

\subsection{Statistical Estimation}\label{section.3.2}
Consider the stochastic stock index process
\[
 \left\{{\bf Y}(t) \vert \,t\ge 0 \right\}=\left\{\left(Y_1(t), \cdots, Y_d(t)\right) \vert \,t\ge 0 \right\},
\]
whose components represent the representative stock index of each country.
We also consider the stochastic currency process
\[
 \left\{{\bf C}(t) \vert \,t\ge 0 \right\}=\left\{\left(C_1(t), \cdots, C_d(t)\right) \vert \,t\ge 0 \right\},
\]
whose components represent the currency of each country in dollar units. 
We define the dollar calibrated stock index, ${\bf X}(t)$, by the currency rate as
\[
 {\bf X}(t):={\bf Y}(t)\cdot {\bf C}(t),
\]
and define the weight vector, ${\bf w}(t)=(w_1(t), \cdots, w_d(t))$, so that the product of the weight and the dollar calibrated index becomes the aggregated value of the stocks as follow
\begin{equation}\label{weight.eq.1}
 {\bf w}(t)\cdot {\bf X}(t) = {\bf V}(t),
\end{equation}
where ${\bf V}(t)=(V_1(t), \cdots, V_d(t))$ is a vector of the aggregated value of the stocks in US dollars at each country at time $t$. 
For the simplicity we assume the weight vector to be a constant through the observed period, hence we denote it as $\bf w$, which enables us to compute 
the aggregated value of the stocks ${\bf V}(t)$ using the dollar calibrated stock ${\bf X}(t)$ and weight $\bf w$ only.  
In this section, we use the final week of April 2013 for the calculation of the weight vector ${\bf w}$ in \eqref{weight.eq.1}.

From \eqref{weight.eq.1}, we can interpret ${\bf w}$ as vector which represent the number of calibrated indices and ${\bf X}(t)$  as a vector which represent the price of the calibrated indices.
Contrary to ${\bf V}(t)$, ${\bf X}(t)$ is the index without unit. Hence, the interpretation in the numbers of $\bf w$ and ${\bf X}(t)$ are somewhat arbitrary. 
However this scale arbitrariness is not an issue in the interpretation of RHIX as explained in Example 2.2..
While this scale arbitrariness leads to a serious interpretation problem in CIX and HIX, the same problem does not cause much problem in the interpretation of RHIX, as discussed in Example 
\ref{example.big} and \ref{example.big.2}.


Previously, we defined RHIX as the summary of herd behavior in an interval $[0,t]$, while we are interested in the spot herd behavior at a fixed time $t=t_0$.
Hence in the following, we define RHIX in a time interval around $t=t_0$, say $[t_0-\epsilon, t_0+\epsilon]$.
For the given time interval
\[
[t_1, t_2]=\left[t_0-\epsilon, t_0+\epsilon\right],
\]
we define RHIX on the interval as follows.
 \begin{equation}\label{11}
  \srhix{[t_1, t_2]}{{\bf w}, {\bf X}}:=
   \frac{\sum\limits_{i\neq j}w_iw_j {\rm cov }[X_i(t_2), X_j(t_2)] }{ \sum\limits_{i\neq j}w_i w_j {\rm cov}[X_i^c(t_2), X_j^c(t_2)] },
 \end{equation}
where $\left\{ {\bf X}(t) \vert t\ge t_1\right\}$ is a stochastic process  whose starting point is  $t=t_1$.

Because RHIX is model free, it can be calculated without any model assumption as did CIX and HIX; see for example, \citet{Skintzi} and \citet{Dhaene2}.
However for the easiness of confidence interval estimation of RHIX, we assume that the stochastic stock process, $\left\{ {\bf X}(t) \vert t\ge t_1\right\}$, follows the multivariate Brownian motion in the given small local interval, $[t_1, t_2]$.
Under the assumption, \eqref{11} can be estimated as follows.
\begin{equation}
   \srhix{[t_1, t_2]}{{\bf w}, {\bf X}}=
   \frac{\sum\limits_{i\neq j}w_iw_j \sigma_{X_i(t_2)}\sigma_{X_j(t_2)}{\rm corr}[X_i(t_2), X_j(t_2)] }{ \sum\limits_{i\neq j}w_i w_j\sigma_{X_i(t_2)} \sigma_{X_j(t_2)} {\rm corr}[X_i^c(t_2), X_j^c(t_2)] },
\end{equation}
where $\sigma_{X_i(t_2)}$ represents
\begin{equation}\label{eq.es.4}
\sigma_{X_i(t_2)}=X_i(t_1)e^{r_i (t_2-t_1)}\sqrt{ e^{\sigma^2_i (t_2-t_1)}-1 },
\end{equation}
and ${\rm corr}[X_i(t_2), X_j(t_2)]$ and ${\rm corr}[X_i^c(t_2), X_j^c(t_2)]$ in the model can be expressed as
\begin{equation}\label{eq.es.2}
{\rm corr}[X_i(t_2), X_j(t_2)]=\frac{e^{\rho_{ij} \sigma_i\sigma_j (t_2-t_1)  }-1}{ \sqrt{ \big(e^{\sigma_i^2 (t_2-t_1)}-1\big) \big(e^{\sigma_j^2 (t_2-t_1)}-1 \big)}  },
\end{equation}
and
\begin{equation}\label{eq.es.3}
{\rm corr}[ X_i^c(t_2), X_j^c(t_2)]=\frac{e^{ \sigma_i\sigma_j  (t_2-t_1)}}{ \sqrt{ \big(e^{\sigma_i^2(t_2-t_1)}-1\big) \big(e^{\sigma_j^2(t_2-t_1)}-1 \big)}   } .
\end{equation}
Here, unknown parameters are $r_{i}$, $\sigma_{i}$ and $\rho_{ij}$ in \eqref{eq.es.4}, \eqref{eq.es.2} and \eqref{eq.es.3}.
In our application, we set $\epsilon=25$ i.e.
\[
 [t_1,t_2]=\left[t_0-\epsilon, t_0+\epsilon \right],
\]
where the unit is a week.
Thus, given the stock prices in the time window
\[
 \left\{ {\bf X}(t_0-\epsilon),\, {\bf X}(t_0-\epsilon+1) \cdots,\, {\bf X}(t_0+\epsilon) \right\},
\]
we have the following i.i.d.  observations
\begin{equation}\label{iid}
 \left\{ {\bf Z}(t_0-\epsilon),\,{\bf Z}(t_0-\epsilon+1),  \cdots,\, {\bf Z}(t_0+\epsilon-1) \right\},
\end{equation}
where  ${\bf Z}(t):=\left( Z_1(t), \cdots, Z_d(t) \right)$ with
\[
 Z_i(t)=\log\frac{X_i(t+1)}{X_i(t)}.
\]
Based on the given i.i.d.  data, the parameter estimates are given as
\[
 \widehat{r}_{i}=\frac{\sum_t  Z_i(t)}{n},\quad \widehat{\sigma}_{i}=\frac{ \left( \sum_t Z_i(t)-\widehat{r}_{i} \right)^2    }{n-1},\quad\hbox{and}\quad \widehat{\rho}_{ij}=
 \frac{    \frac{  \sum_t \left(  Z_i(t)-\widehat{r}_{i} \right)   \left(  Z_j(t)-\widehat{r}_{j} \right)     }{n-1}    }{   \widehat{\sigma}_{i} \widehat{\sigma}_{j}    } ,
\]

where summations are on the set $\{t_0-\epsilon, t_0-\epsilon+1, \cdots, t_0+\epsilon-1 \}$.
Because observations in \eqref{iid} are i.i.d, we can easily estimate the confidence interval of HIX and RHIX using the bootstrap method.
Finally we note that the local multivariate Brownian motion assumption can minimize model risk, see for example \citet{Pauline}.

\subsection{Results}
In this section, the herd behaviors of different continents are compared using the MSCI stock index data. Data from 1996 to Apr. 2013 are used for the main  analysis but data of Africa and the Middle East 
can not be used in the 90's because the exchange rates to the US dollar are specified after 2000. Through this comparative analysis, peculiar features, resemblances, differences, and trends of each 
continent's herd behavior are observed; however this paper focuses on the observed phenomena but not their causes, which goes beyond the scope of this study.

\begin{figure}[!ht]
    \subfloat[ASIA RHIX and Stock Index (Without China)]{%
      \includegraphics[width=0.49\textwidth]{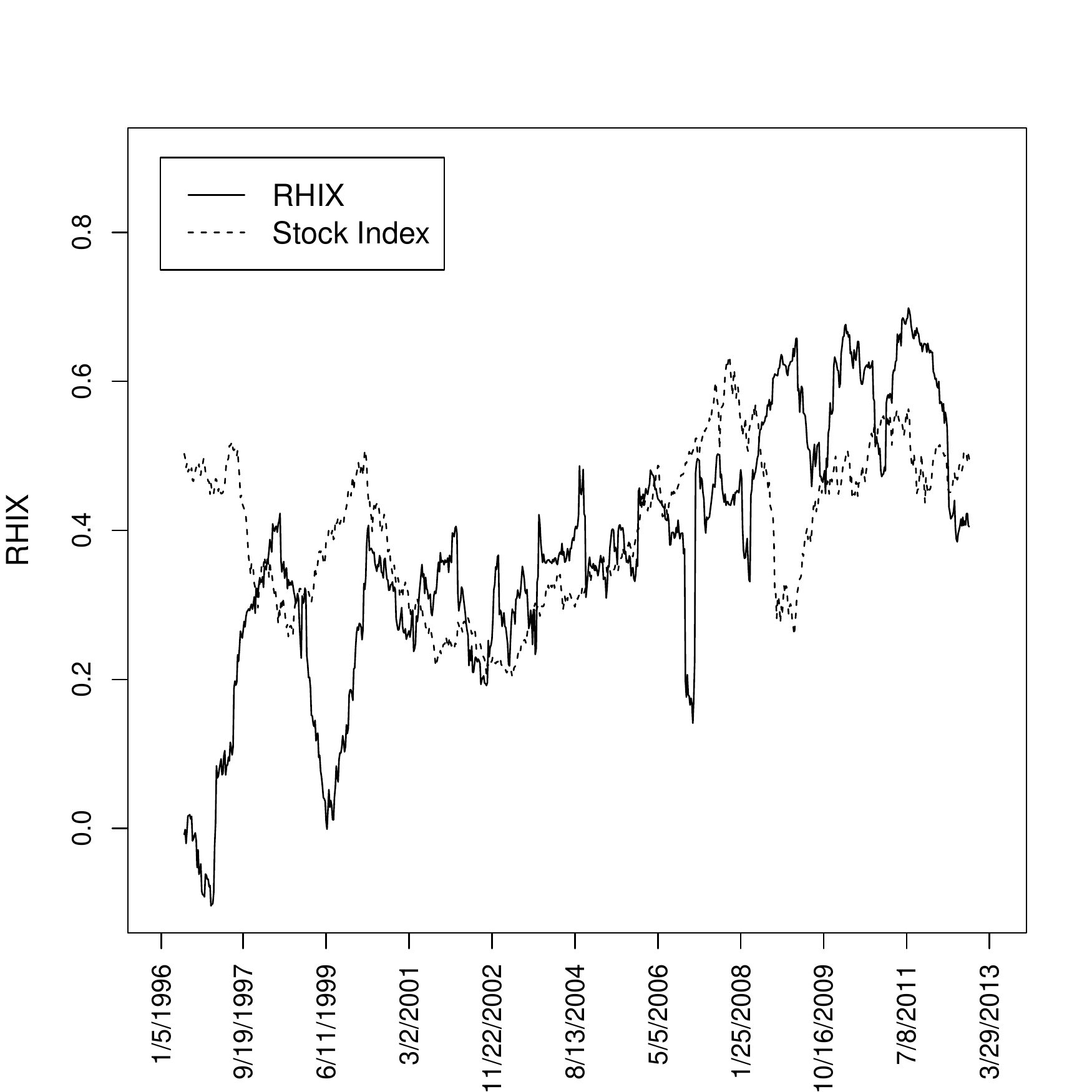}
    }
    \hfill
    \subfloat[ASIA RHIX (With and Without China)]{%
      \includegraphics[width=0.49\textwidth]{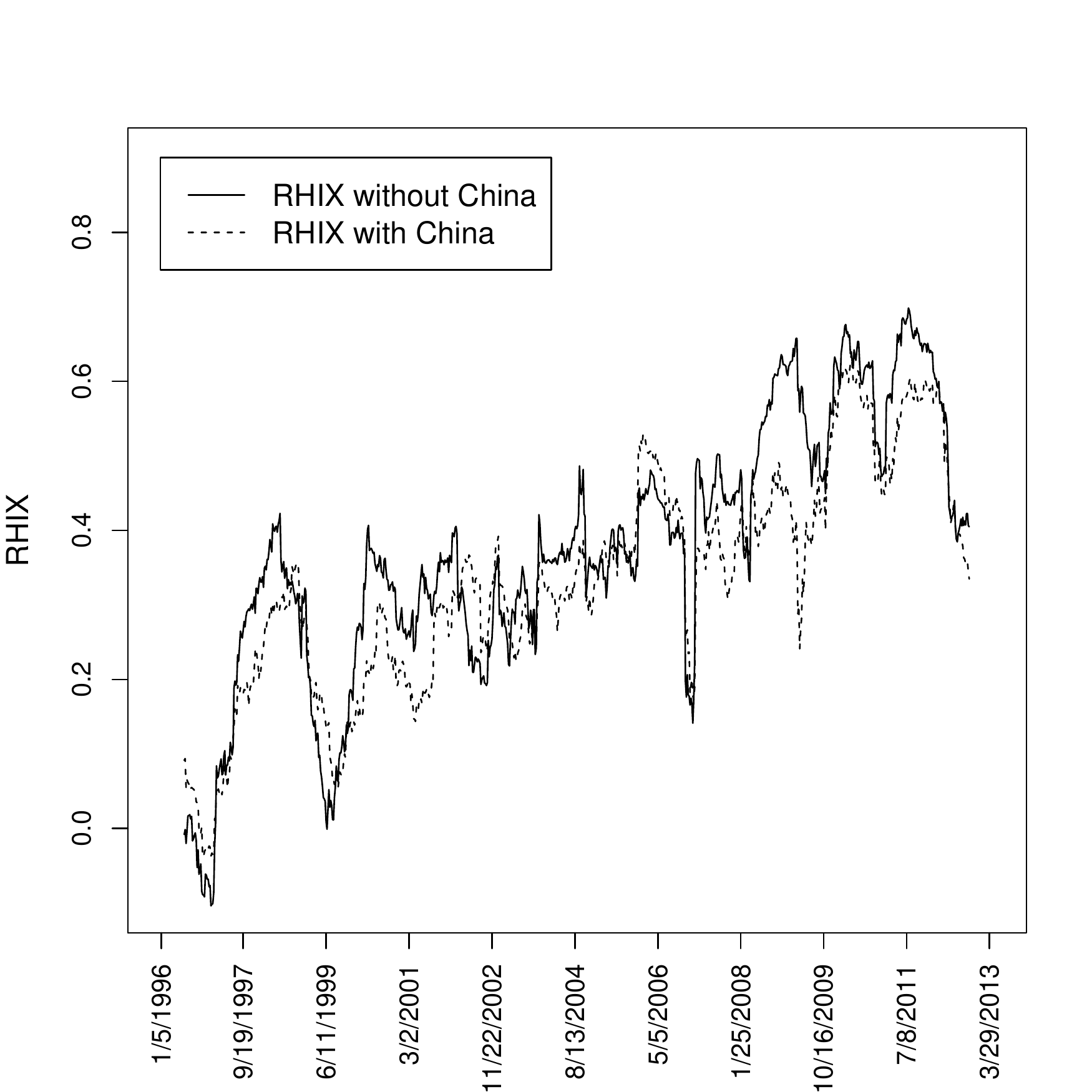}
    }
      \caption{ASIA RHIX}
     \label{ASIA RHIX}
  \end{figure}

\begin{figure}[!ht]
  \centering
  \subfloat[Europe RHIX and Stock Index]{\includegraphics[width=.49\textwidth]{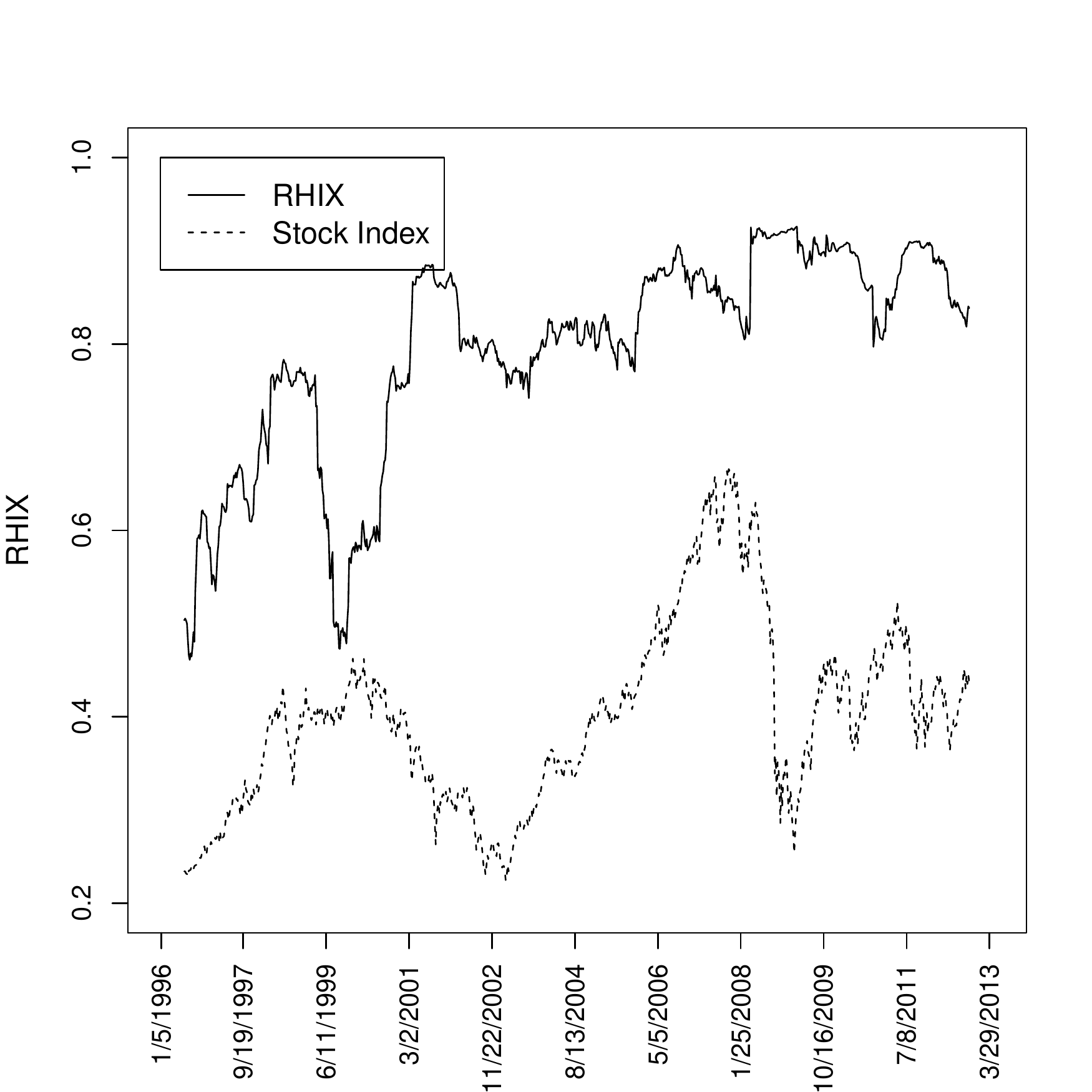}}
  \subfloat[Africa RHIX and Stock Index]{\includegraphics[width=.49\textwidth]{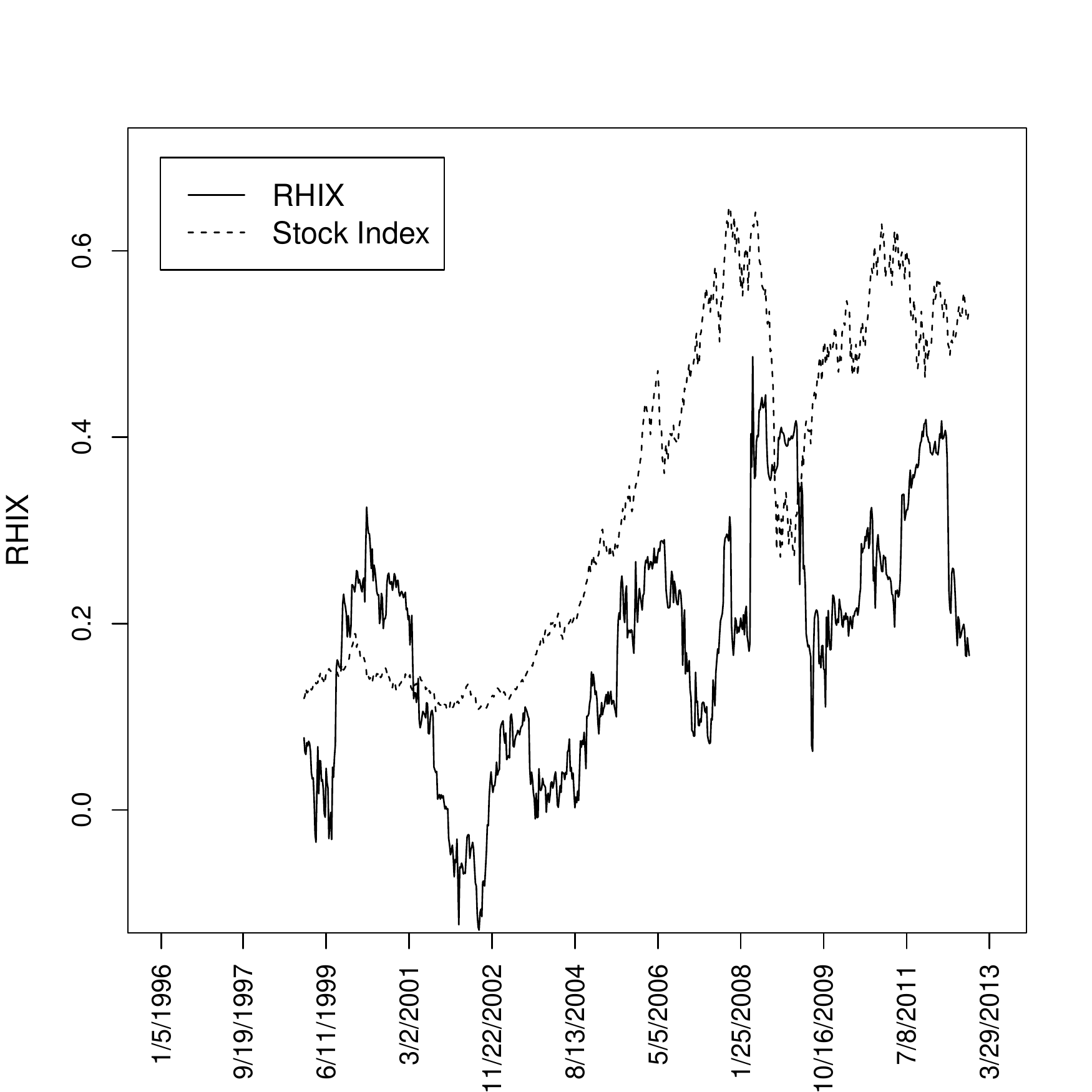}}\\
  \subfloat[The Middle East RHIX and Stock Index]{\includegraphics[width=.49\textwidth]{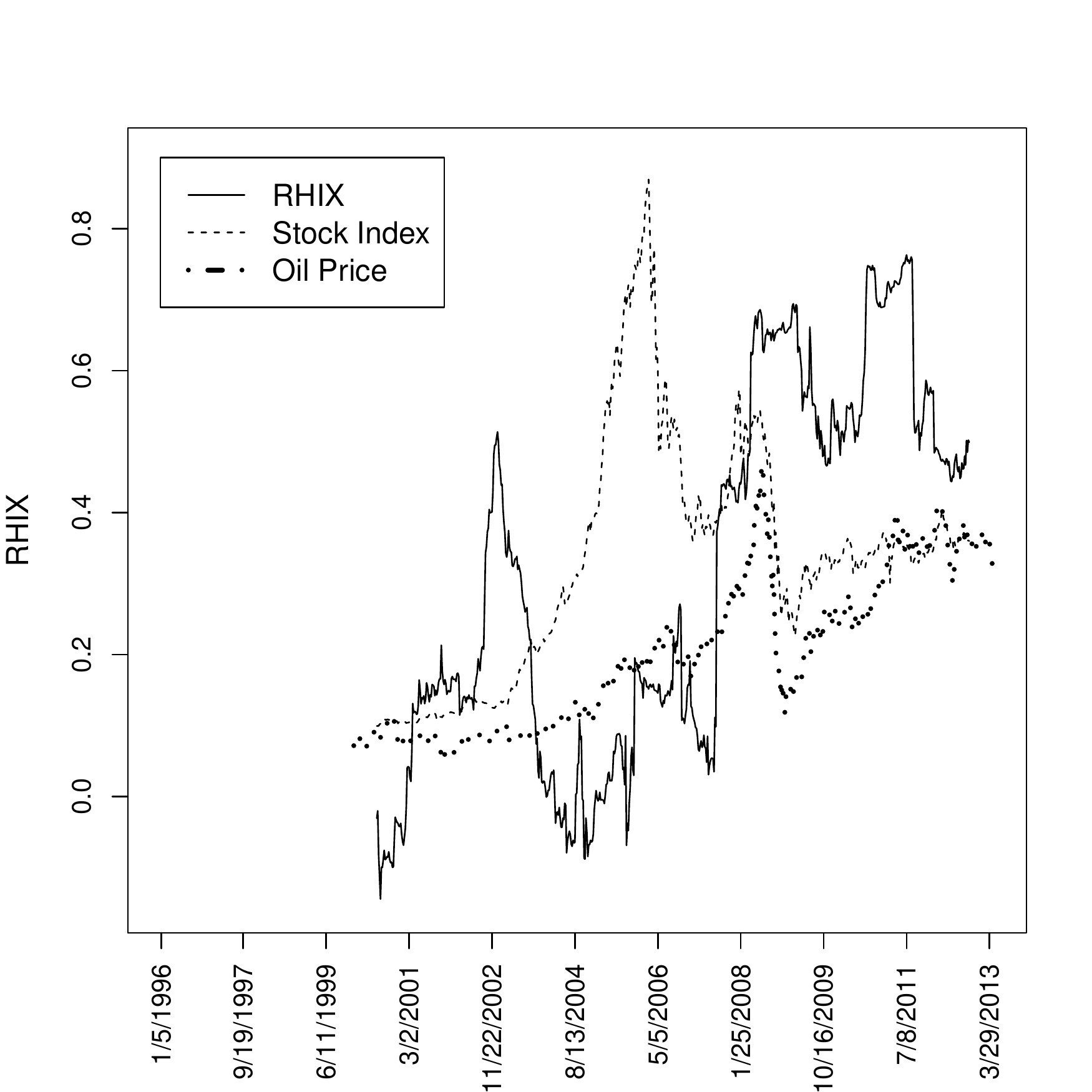}}
  \subfloat[South America RHIX and Stock Index]{\includegraphics[width=.49\textwidth]{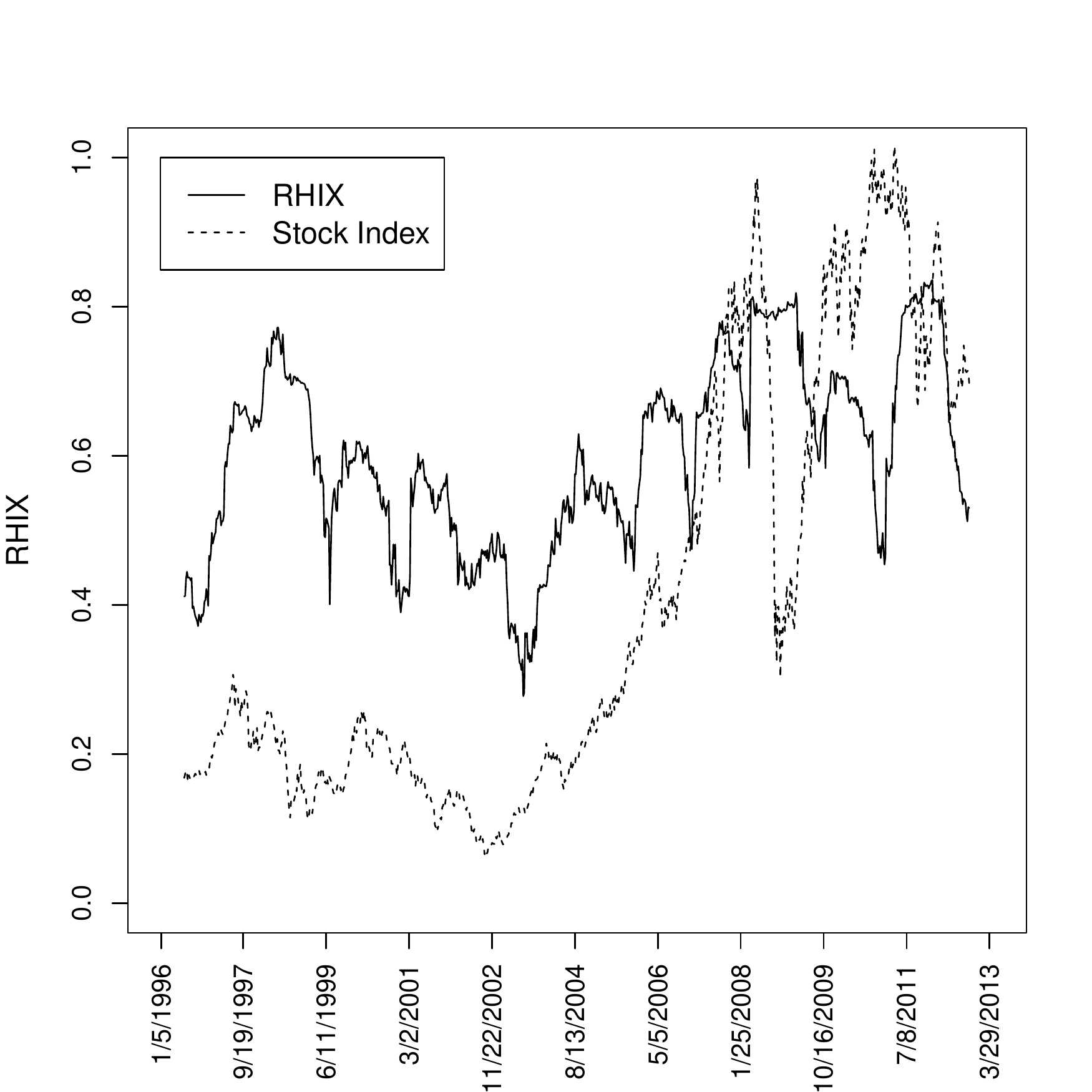}}
  \caption{RHIX and Stock Index: Europe, Africa, the Middle East, and South America}
  \label{fig:sub1}
  \end{figure}

\begin{figure}
\includegraphics[width=0.6\linewidth]{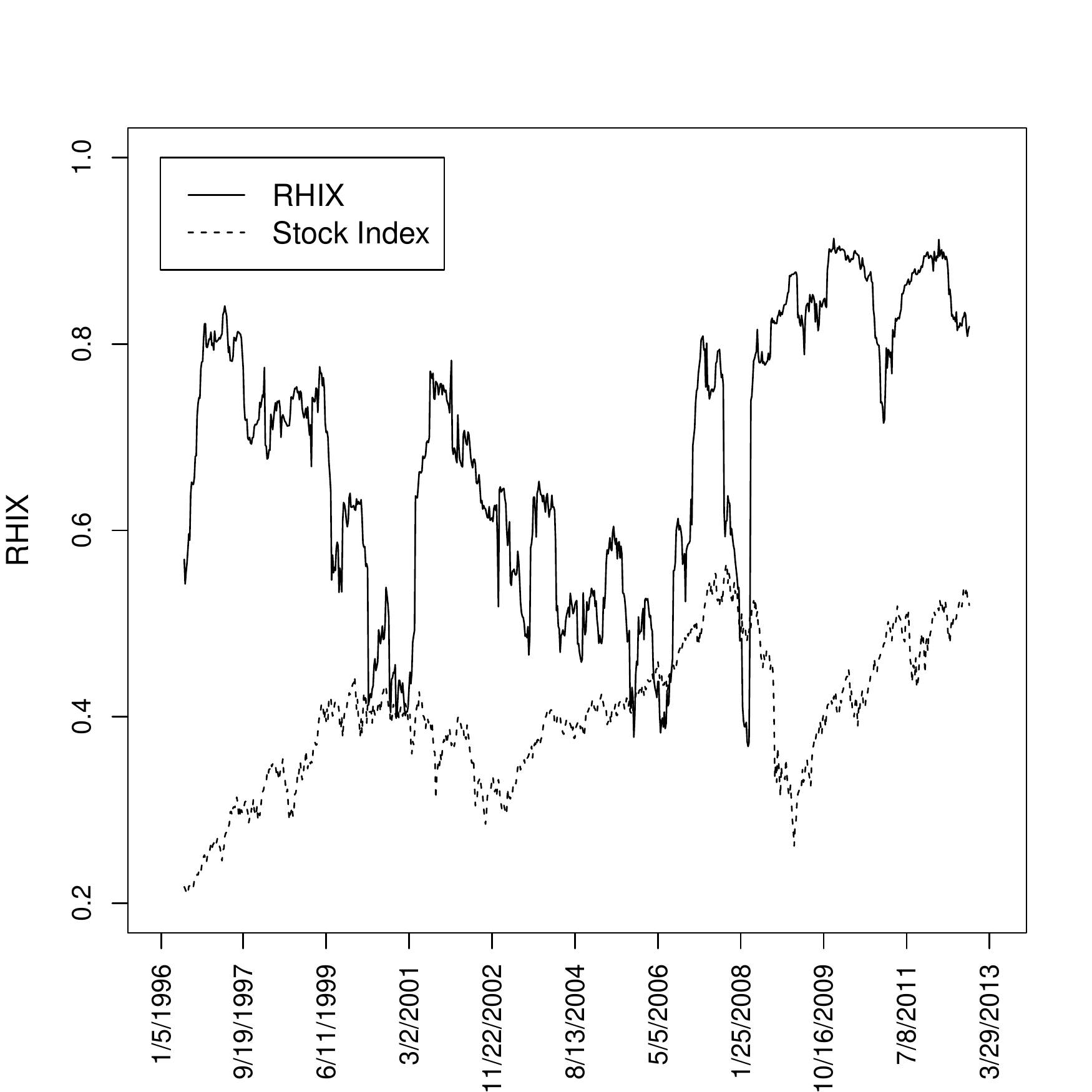}
         \centering\caption{North America RHIX and Stock Index}\label{fig:northamerica}
\end{figure}

A comparative analysis on the herd behavior from each continent at various financial crises is conducted as follows. The RHIX of each continent is described in Figures \ref{ASIA RHIX},
\ref{fig:sub1}, and \ref{fig:northamerica}.

Firstly, during the period of the Asian financial crisis, Asia's RHIX surged rapidly and that of South America also increased. This coincides with \citet{Scholes}, which
describes the effect of the Asian financial crisis on South America. No such increase has been observed in other continents' RHIXs, as the Asian financial crisis mainly impacted regions of Asia and South America.

Next, herd behavior due to the dot-com bubble burst that began in 2000 is analyzed. The RHIXs of most continents revealed a gradual increase during the period,
while there were rapid drops in the stock indices during the same period.

The analysis during the period from 2001 to 2005 reflects the occurrence of the September 11 attacks and the Iraq War. As specified in \citet{Nanto}, the world experienced a small recession immediately after 9/11, and this was reflected in the increase of the RHIX in most continents. Throughout the period, a sharp rise in the RHIXs of the Middle East and North America was shown; other RHIXs experienced small increases except for that of Europe, which decreased.

RHIXs are also analyzed during the period from 2008 to the present in order to consider the global financial crisis and the European sovereign debt crisis. During the global financial crisis
period, all continents showed the maximum RHIX. Until today, most of the continents maintain a fairly high RHIX, which indicates the severe and persistent impact of the global financial crisis and the European sovereign debt crisis on all continents.

As shown in Figure \ref{ASIA RHIX}, the RHIXs of Asia during and after the Asian and global financial crises have an increasing trend. 
In fact, our observation is coherent with \citet{ChengH} and \citet{Nieh}, where the herd behavior of the stock market in Asia was described during and after the Asian financial crisis and the global financial crisis.
As shown in Figure \ref{ASIA RHIX} (b), it is interesting to observe that 
the RHIX of Asia is reduced when China is considered in the analysis. Specifically, the effect of China decreased the level of the RHIX of Asia substantially during the recent 
global financial crisis and the European sovereign debt crisis periods. This observation is also found in \cite{Morrison}, which describes the emergence of the Chinese stock market during the recent global financial crisis.

We can observe the main economic crises that affected Europe from Figure \ref{fig:sub1} (a). The first is the economic crisis that commenced in early 2000. This is related to the dot com bubble burst starting in 2000. The second is the global financial crisis and the European sovereign debt crisis. During the crisis periods, stock prices dropped sharply and RHIX soared rapidly.
It appears that the Asian financial crisis of 1997 did not significantly affect the European economy or herd behavior. The European RHIX is very high for the entire observation period.
This is because European economies are connected under the EU's economic and monetary union. From 2002 to 2008, the stock market index increased, whereas the RHIX stayed nearly constant at around 0.7. Over the entire observation period, 
the RHIX in Europe slowly increased. The next referential papers explain the trend of RHIX shown in Figure \ref{fig:sub1} (a). \citet{Horta} analyze the change of the stock index in Europe before and after the global financial crisis using copula. 
After the crisis, the financial contagion in the stock market is revealed and comovements between the analyzed stock markets have become more noticeable. \citet{Baglioni} and \citet{Metiu} prove that the financially cross-border contagion in Europe increases on and after the sovereign debt crisis.

The African RHIX is around 0 from 1997 to 2005 except for the years 1999 and 2000, as shown in Figure \ref{fig:sub1} (b). 
We can observe that the RHIX of Africa is high during the global financial crisis, which is coherent with \citet{Abdul} 
where the relation between the global financial crisis and some developing countries in Africa, such as Nigeria is reported. 
However, even during the global financial crisis, the RHIX was around $0.4$, which is relatively low compared to RHIX of other continents. 


The RHIX in the Middle East was as low as 0 from 2000 through 2001 and from 2003 to 2005, as shown in Figure \ref{fig:sub1} (c). It then became very high around 9/11, the
Iraq War, the global financial crisis, and the European sovereign debt crisis. Along with the highly volatile stock index, the Middle East's RHIX also demonstrated the most dramatic changes compared to those for other continents. The high RHIX after 9/11 indicates that herd behavior commenced among countries in the Middle East. It is interesting to observe that the RHIX was below
$0$, when the stock price increased rapidly from 2003 to 2005. According to \citet{Rjoub}, stock market returns in the Middle East move negatively most of time during the global financial crisis. However, stock returns temporarily increase, particularly in Jordan, during the Iraq War because money transfers from Iraq to Jordan.

South America's RHIX was very high during the Asian financial crisis and the global financial crisis, as shown in Figure \ref{fig:sub1} (d). 
Further stock prices in South America during the observation period are more dynamic compared to that for other continents. Such high herd behavior and dynamic movement of stock market in 
South America are also described in \citet{Almeida} and \citet{Salman}.

From 2001 through to 2006, the RHIX in North America decreased and the stock price increased, as shown in Figure \ref{fig:northamerica}. 
Although this is reasonable, it cannot be observed in other continents. In the calculation of the RHIX, Mexico is almost excluded because its aggregate stock prices are too low. Hence, we can, in essence, regard that Canada has been highly correlated to the USA. This explains why the RHIX is so high for almost the entire observation period for North America. 
Even during the Asian financial crisis, while the stock price was increasing, the RHIX was very high. 

\begin{figure}[!ht]
  \centering
  \subfloat[Comparison of RHIX: Europe, Asia, and Africa]{\includegraphics[width=.49\textwidth]{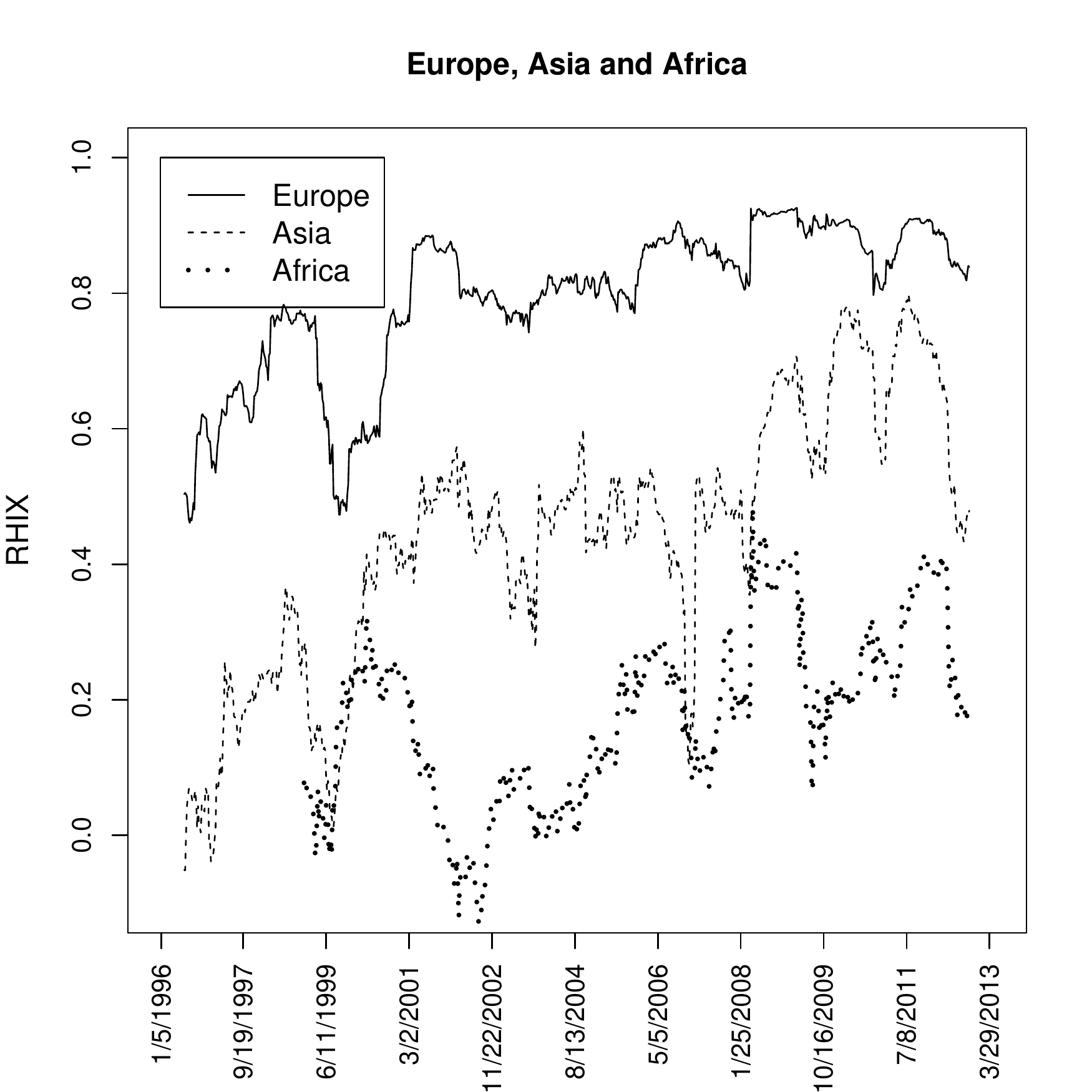}}
    \subfloat[Comparison of RHIX: The Middle East, Africa, and Asia]{\includegraphics[width=.49\textwidth]{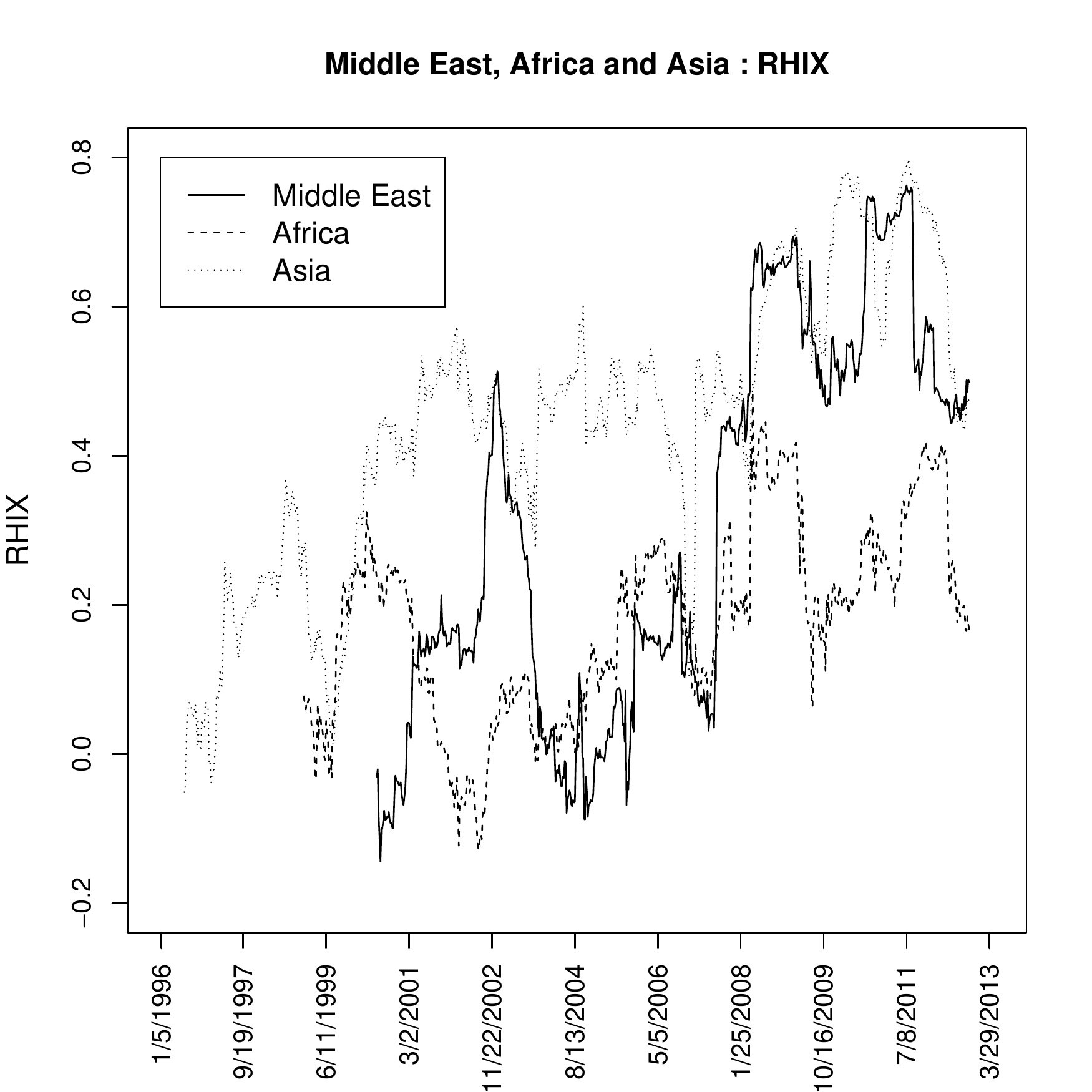}}\\
  \subfloat[Comparison of Stock Prices: Europe, Asia, and Africa]{\includegraphics[width=.49\textwidth]{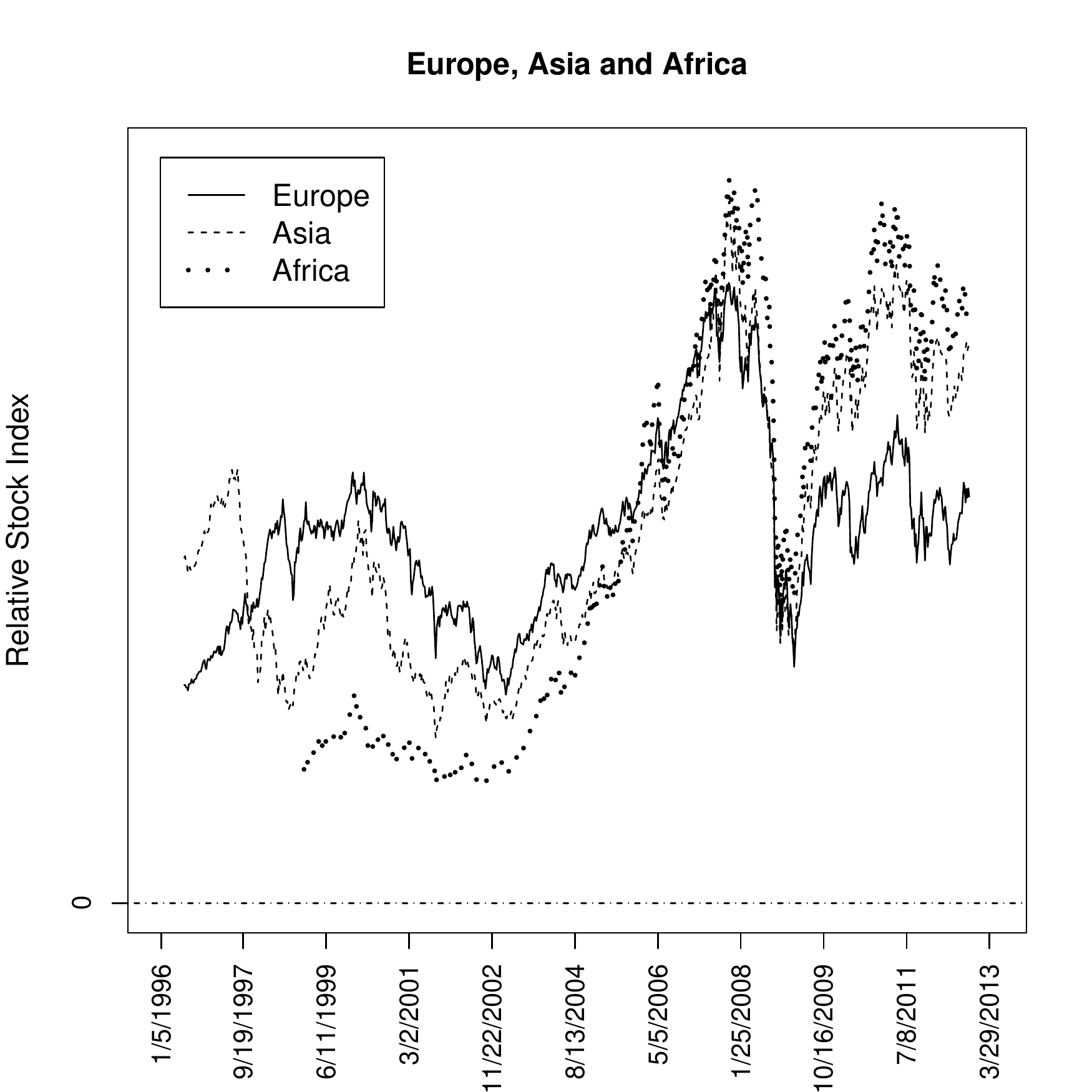}}
  \subfloat[Comparison of Stock Prices: The Middle East, Africa, and Asia]{\includegraphics[width=.49\textwidth]{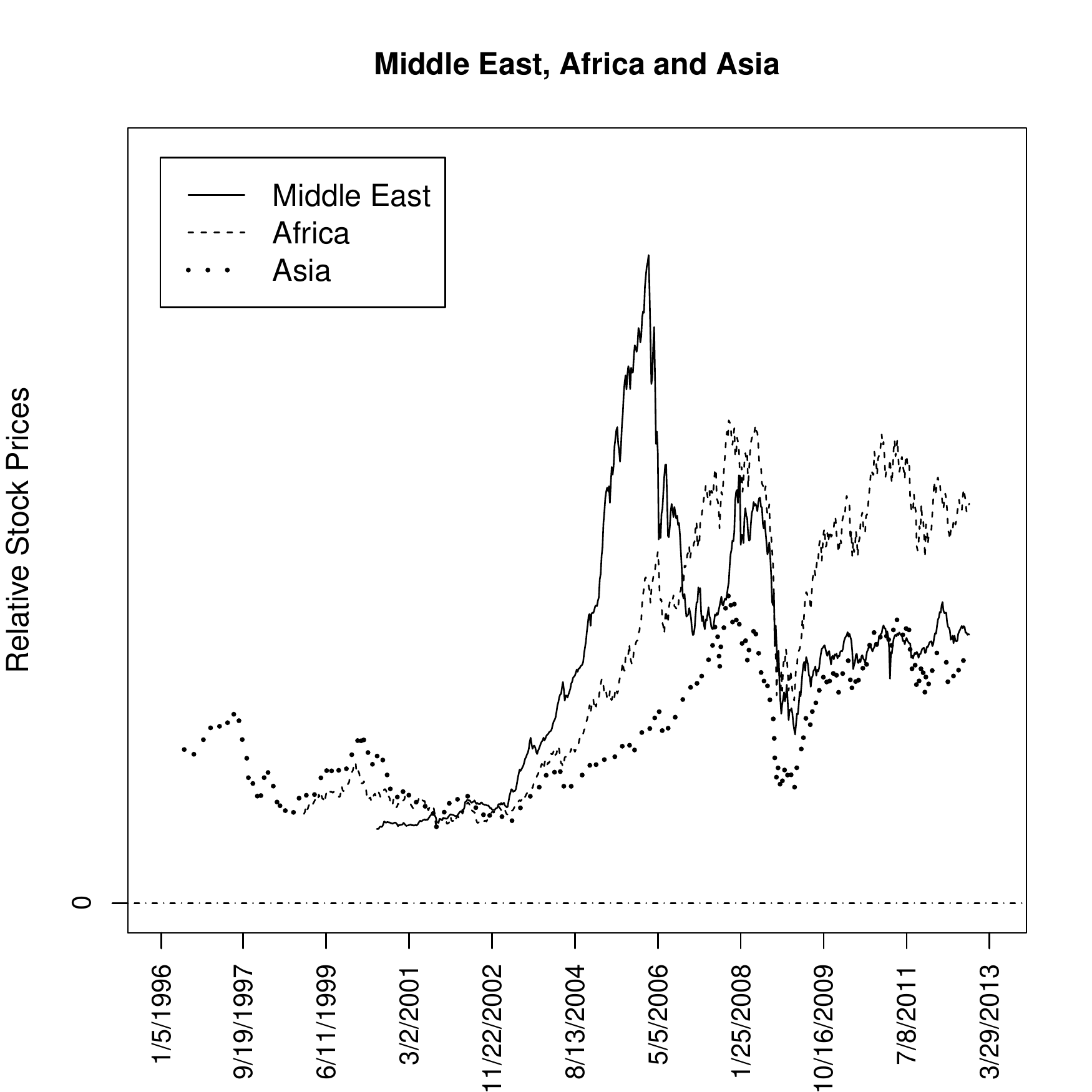}}
  \caption{Comparison of RHIXs and stock prices among continents}
  \label{fig:sub2}
  \end{figure}

From another point of view, we carry out a further comparative analysis of some continents. 
In Europe, Asia, and Africa, the RHIXs were at their maximum during the global financial crisis and at their local maximum during the European sovereign debt crisis, 
as shown in Figure \ref{fig:sub2} (a). However, the RHIX in Asia is unusually high; yet this is not the case for Europe during the Asian financial crisis. 
As Figure \ref{fig:sub2} (c) illustrates, the stock price movement pattern in the three continents is somewhat similar after 2004. During the Asian financial crisis, Asian markets demonstrated a dramatic volatility.

As Figure \ref{fig:sub2} (b) shows, the RHIX of the Middle East is as low as Africa's when it is at a lower stage. However, when it is at a higher stage, it exceeds that of Asia. This is indicative of the extreme movement of herd behavior in the Middle East, 
which experiences instability due to oil prices and politics. Figure \ref{fig:sub2} (d) also shows the extreme dynamics of stock prices in the Middle East compared to those of Africa and Asia.

Generally, the order of the RHIX scale is as follows. The RHIXs in Europe and North America maintain the highest level, followed by, in order, South America, Asia, the Middle East, and Africa.
Nevertheless, the RHIX shows a similar pattern among the continents in the recent period and an increasing trend over
time. This is related to the globalization of the world economy. Table \ref{table10} shows the RHIXs of all continents in special periods.

\begin{figure}[!ht]
    \subfloat[RHIX in Africa: $95\%$ Confidence Interval]{%
      \includegraphics[width=0.49\textwidth]{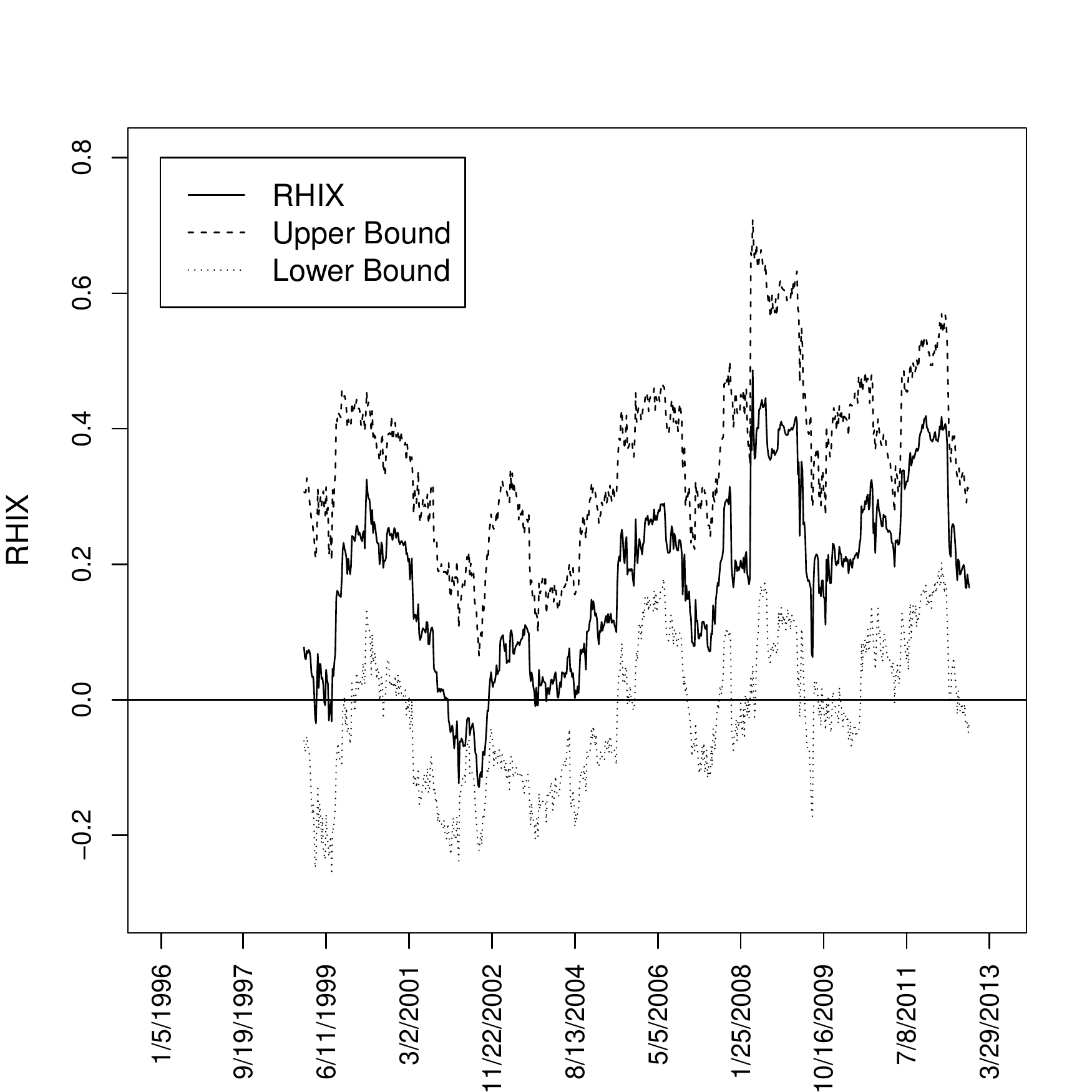}
    }
    \hfill
    \subfloat[RHIX in the Middle East: $95\%$ Confidence Interval]{%
      \includegraphics[width=0.49\textwidth]{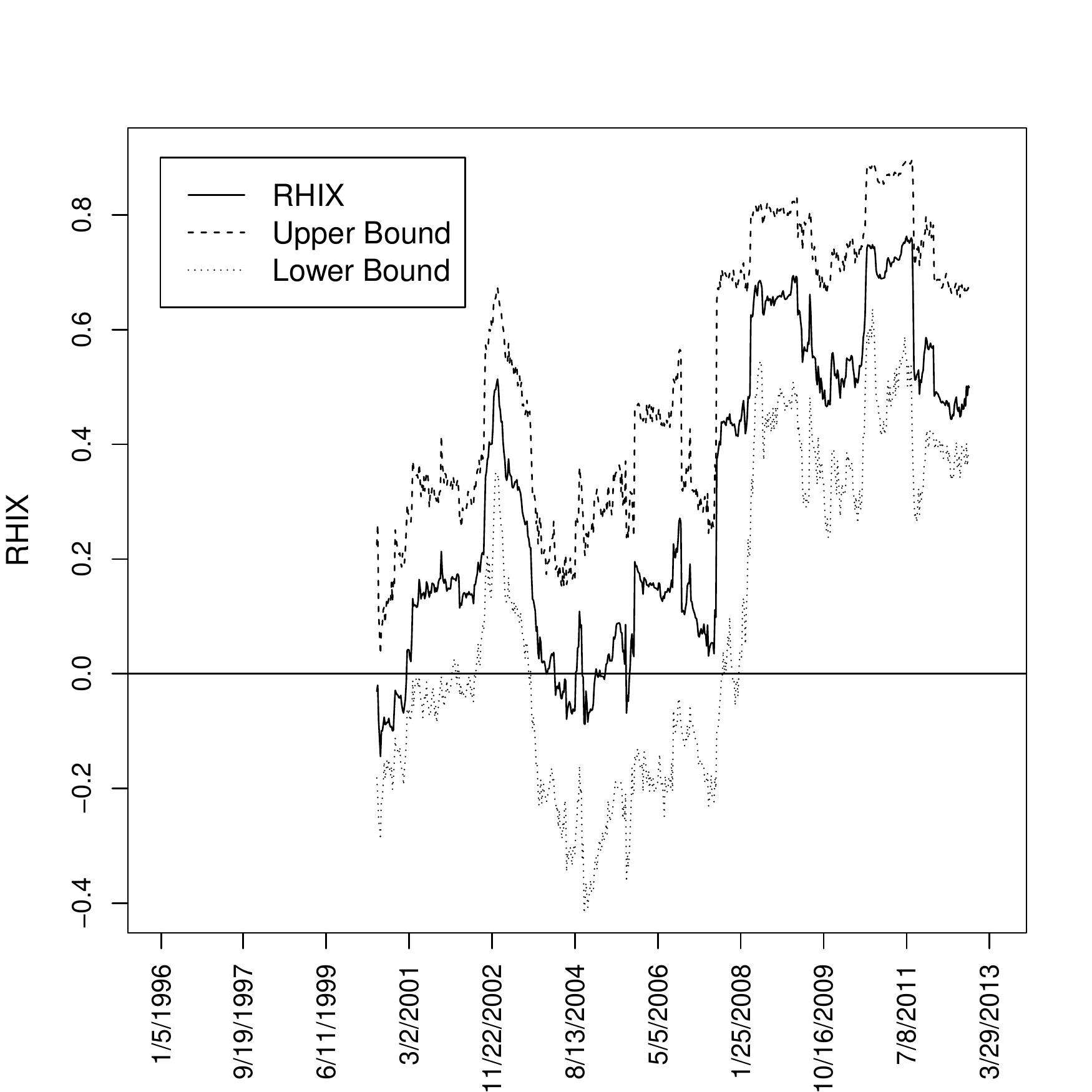}
    }
      \caption{Confidence Interval of the RHIX with Resample Size $1,000$}
     \label{confidence}
  \end{figure}

Finally, we briefly mention the confidence interval of the RHIX. As specified earlier, the RHIX of Africa and the Middle East are not only relatively smaller than that of the other continents,
but also close to $0$ in many time intervals. Hence, it is interesting to investigate the confidence interval of the RHIX in the two continents. For the confidence interval estimation, we use
the bootstrap method with resample size $1,000$. As seen from Figure \ref{confidence}, in most time interval, the  $95\%$ confidence intervals of RHIXs include $0$. 

 \begin{table}
 \begin{tabular}{| c | c | c | c | c |}
 \hline
    Continents & \multicolumn{4}{|c|}{Avereage RHIX(Standard Deviation)}  \\
    \cline{2-5}
     & AFC & DBB & GFC & Entire Periods   \\
 \hline
 Asia & 0.3153(0.0589) & 0.3157(0.0436) & 0.9581(0.0633) & 0.3723(0.1732)   \\
 Europe & 0.7117(0.0572) & 0.7236(0.1043) & 0.8906(0.0309) & 0.7932(0.1116) \\
 North America & 0.7346(0.0351) & 0.4983(0.0879) & 0.8524(0.0408) & 0.6818(0.1467)  \\
 South America& 0.6937(0.0397) & 0.5242(0.0716) & 0.7019(0.0957) & 0.6000(0.1252)\\
 Africa & NA & 0.2140(0.0570) & 0.2814(0.0850) & 0.1770(0.1290)\\
 The Middle East& NA & $-0.0036(0.0921)$ & 0.6159(0.0911) & 0.3090(0.2589)\\
  \hline
\end{tabular}

 \caption{Summarization of RHIX in different periods}
\begin{tablenotes}
      \small
      \item \emph{Notes}: AFC(Asian financial crisis), DBB(dot com bubble burst), GFC(global financial crisis), and NA(not available)
\end{tablenotes}
\label{table10}
 \end{table}

\section{Conclusion}
Most of the previous studies in this field have examined empirical evidence on herd behavior in a country.
In this case, because the herd behavior is interpreted by comparing the current degree of comonotonic behavior with that in the past, the CIX or HIX can be applied.
However, we note that the comparison between two HIXs, for example, is not easy because HIXs with different weights and volatilities have different measurement bases.
As pointed out, CSSD is not useful for a comparative study neither. Thus, as an alternate measure of herd behavior, we have proposed RHIX and exploited its main properties using empirical analysis.

In this paper, we investigated herd behavior within each continent and among continents.
In particular, the latter study required a comparison of the herd behavior in one continent relative that in others, and therefore the use of
RHIX is appropriate. Using RHIX, both well-known facts and new interesting facts are observed through empirical analysis.
During the initial period of our empirical analysis, herd behavior and the stock index were negatively correlated, i.e.,
the RHIX was high when the stock index dropped rapidly, while the RHIX was relatively low when the stock index was increasing.
However, this pattern has weakened over time as financial globalization has increased. The RHIX also shows an increasing trend across all of the continents, but the slopes differ significantly. It is noteworthy that major financial crises such as the global financial crisis in 2008 and the European sovereign debt crisis affected all the countries significantly and led to high RHIXs in most continents.

We demonstrate the period of strong herding. For example, the RHIXs from most of the continents hit the upper limit during the period of the global financial crisis.
However, only Asia and South America showed a significant degree of comovement during the period of the Asian financial crisis.
We also observe that the RHIXs of all continents have been increasing with time. This is a strong evidence that the world forms a financially unified body.
Furthermore, generally, higher RHIXs are observed in developed countries than in developing countries. The level of information delivery might be one of the reasons for the higher RHIXs in developed countries; 
however, more research should be done on this phenomenon.

Although the current model is believed to capture the essential characteristics of herd behavior,
more elaborate stock index models can be considered. This topic should be further analyzed in order to see how RHIX depends on
the choice of these models.
It is also interesting to apply RHIX to the stock index data with different time scales.
We may expect herd behavior measured on a daily basis to differ from that measured on a weekly basis.
Currently, we use a weekly-basis stock index, so herd behavior on different time scales should be addressed with caution.

\appendix

\section*{Appendix}%
\noindent\textbf{Definition A.1}(\citet{Dhaene}).
Let ${\bf x}=(x_1,\cdots,x_d)\in \Real^d$ and ${\bf y}=(y_1,\cdots, y_d)\in \Real^d$. A set $A\subset R^d$ is comonotonic if the following inequality holds
 \[
  (x_1-y_1)(x_2-y_2)\cdots(x_d-y_d)\ge 0 \quad\hbox{for all}\quad {\bf x},\,{\bf y}\in A.
 \]
We also denote $X$ as comonotonic if it has comonotonic support, i.e. if $P(X\in A)=1 $ for some comonotonic set $A\subset \Real^d$.

\noindent\textbf{Theorem A.1}(\citet{Dhaene}).
 The following statements are equivalent.
 \begin{enumerate}
  \item[1.] $X$ is comonotonic
  \item[2.] $P({\bf X}\le {\bf x})=\min\left\{F_1(x_1), \cdots, F_d(x_d) \right\}$
  \item[3.] For $X\sim \hbox{Uniform\rm($0,1$)}$, we have
  \begin{equation}
   {\bf X}\eqd \left(F_{X_1(t)}^{-1}(U), \cdots, F_{X_d(t)}^{-1}(U) \right),
  \end{equation}
where $\eqd$ means equal in distribution.
 \end{enumerate}

\noindent\textbf{Definition A.2}(\citet{Dhaene2}).
 The following statements are equivalent
\begin{itemize}
 \item[1.] $(X_1,\cdots, X_d)$ is comonotonic
 \item[2.] $S\eqd S^c$
 \item[3.] ${\rm Var}[S(t)] = {\rm Var}[S^c(t)]$
\end{itemize}


\noindent\textbf{Proof of Proposition} \textbf{\ref{prop.cix}.}
First observe that
\[
 0\le {\rm var}\left[S(t)\right]\le {\rm var}\left[S^c(t)\right],
\]
where the first inequality is trivial with equality holds if and only if
\[
 P\left(\sum\limits_{i=1}^d w_i X_i(t)=c\right)=1.
\]
Further the second inequality is from Theorem 6 of \citet{Cheung3} with equality holds if and only if ${\bf X}(t)$ is comonotonic.

Because the numerator of CIX in \eqref{def.cix} can be expressed as
\begin{equation}\label{cix.proof.1}
 \sum\limits_{i\neq j} w_i w_j {\rm cov}\left( X_i(t), X_j(t) \right) = {\rm var}\left[ S(t) \right] -\sum\limits_{i=1}^{d}w_i^2 {\rm var}\left[ X_i(t) \right],
\end{equation}
the proofs of 1., 2. and 3. are immediate from the above observations and \eqref{cix.proof.1}. The proof of 4. is obvious from the definition of CIX.
 \hfill$\Box$

\noindent\textbf{Proof of Proposition} \textbf{\ref{prop.rhix}.}
First note that the following condition
\[
 P\left(\sum\limits_{i=1}^d w_i X_i(t)=c\right)=1
\]
implies
\begin{equation}\label{c}
 c=\sum\limits_{i=1}^d w_i \mu_{X_i(t)}.
\end{equation}
Then for the given $c$ as in \eqref{c}, we have the following inequality
\begin{equation*}
 \begin{aligned}
  \E{ \left( \sum\limits_{i=1}^{d}w_iX_i(t)-c \right)^2 }&=\E{\sum\limits_{i=1}^{d}w_i^2X_i(t)^2 +\sum\limits_{i\neq j}w_iw_jX_i(t)X_j(t) +c^2 -2c\sum\limits_{i=1}^{d}w_iX_i(t) }\\
  &\ge 0
 \end{aligned},
\end{equation*}
which is equivalent to
\begin{equation}\label{proof.prop.2}
 \begin{aligned}
  \E{  \sum\limits_{i\neq j}w_iw_jX_i(t)X_j(t) } &\ge  2c\sum\limits_{i=1}^{d}w_i\mu_{X_i(t)} - c^2 -\sum\limits_{i=1}^{d}w_i^2(\sigma_{X_i(t)}^2+\mu_{X_i(t)}^2) \\
  &= c^2 - \sum\limits_{i=1}^{d}w_i^2(\sigma_{X_i(t)}^2+\mu_{X_i(t)}^2)\\
  &=\sum_{i\neq j} w_iw_j \mu_{X_i(t)} \mu_{X_j(t)} - \sum\limits_{i=1}^{d}w_i^2\sigma_{X_i(t)}^2,
 \end{aligned}
\end{equation}
where the equality in the first inequality in \eqref{proof.prop.2} holds if and only if
\[
 P\left(\sum\limits_{i=1}^d w_i X_i(t)=c\right)=1.
\]
Part 1 is an immediate result from \eqref{proof.prop.2}. Finally, 2 is an easy application of Theorem {A.2}, and 3 is trivial from the definition of RHIX.
 \hfill$\Box$

\noindent\textbf{Proof of Proposition} \textbf{\ref{prop.linear.1}.}
Since the proof of HIX is similar to that of RHIX, we only prove RHIX here.
Now we prove the first part.
For convenience, define
 \[
 \begin{aligned}
  {\bf G}(t)&:=\alpha{\bf X}(t)+{\bf b}\\
  &=(\alpha X_1(t)+b_1, \cdots, \alpha X_d(t)+b_d).
 \end{aligned}
 \]
 From equations
 \[
  {\rm cov}\left[G_i(t), G_j(t) \right]={\rm cov}\left[\alpha X_i(t), \alpha X_j(t) \right] \quad \hbox{and} \quad {\rm cov}\left[G_i^c(t), G_j^c(t) \right]={\rm cov}\left[\alpha X_i^c(t), \alpha X_j^c(t) \right],
 \]
we can conclude that
\[
 \rhix{{\bf w}, {\bf G}(t)}=\rhix{{\bf w}, {\bf X}(t)}.
\]
Finally, from
 \begin{equation}
 \begin{aligned}
 {\rm cov}_{{\bf a}\cdot {\bf w}}\left[ {\bf a}^{-1} \cdot{\bf X}(t) \right]  &= \sum\limits_{i\neq j} a_iw_ia_jw_j {\rm cov}\left[ \frac{X_i(t)}{a_i}, \frac{X_j(t)}{a_j}\right]\\
 &={\rm cov}_{{\bf w}}\left[ {\bf X}(t)\right],
 \end{aligned}
 \end{equation}
 the proof  of the second part is immediate.  \hfill$\Box$

\begin{center}
{\bf References} 
\end{center}

\bibliographystyle{apalike}
\bibliography{CTE_Bib_HIX}
\end{document}